\documentclass[aps,twocolumn]{revtex4}

\RequirePackage[T2A]{fontenc}

\usepackage{amssymb}
\usepackage{amsmath}
\usepackage{bm}
\usepackage[dvips]{epsfig}
\usepackage{graphicx}

\begin{document}

\title{Distinctive features of oscillatory phenomena in 
reconstructions of the topological structure of electron 
trajectories on complex Fermi surfaces}

\author{A.Ya. Maltsev}

\affiliation{
\centerline{\it Steklov Mathematical Institute}
\centerline{\it Russian Academy of Sciences}
\centerline{\it Moscow, 119991 Russia}
}

\begin{abstract}
We consider the behavior of classical and quantum oscillations 
in metals with complex Fermi surfaces near the directions of
$\, {\bf B} \, $ corresponding to changes in the topological 
structure of the dynamical system describing the semiclassical 
motion of quasiparticles along the Fermi surface. The transitions 
through the boundaries of change in this structure are accompanied 
by sharp changes in the picture of oscillations, the form of which 
depends in the most essential way on the topological type of the 
corresponding reconstruction. We list here the main features of 
such changes for all topological types of elementary reconstructions 
and discuss the possibilities of experimental identification of such 
types based on these features. 
\end{abstract}

\maketitle

\vspace{5mm}

\section{Introduction}

\begin{figure}[t]
\begin{center}
\includegraphics[width=\linewidth]{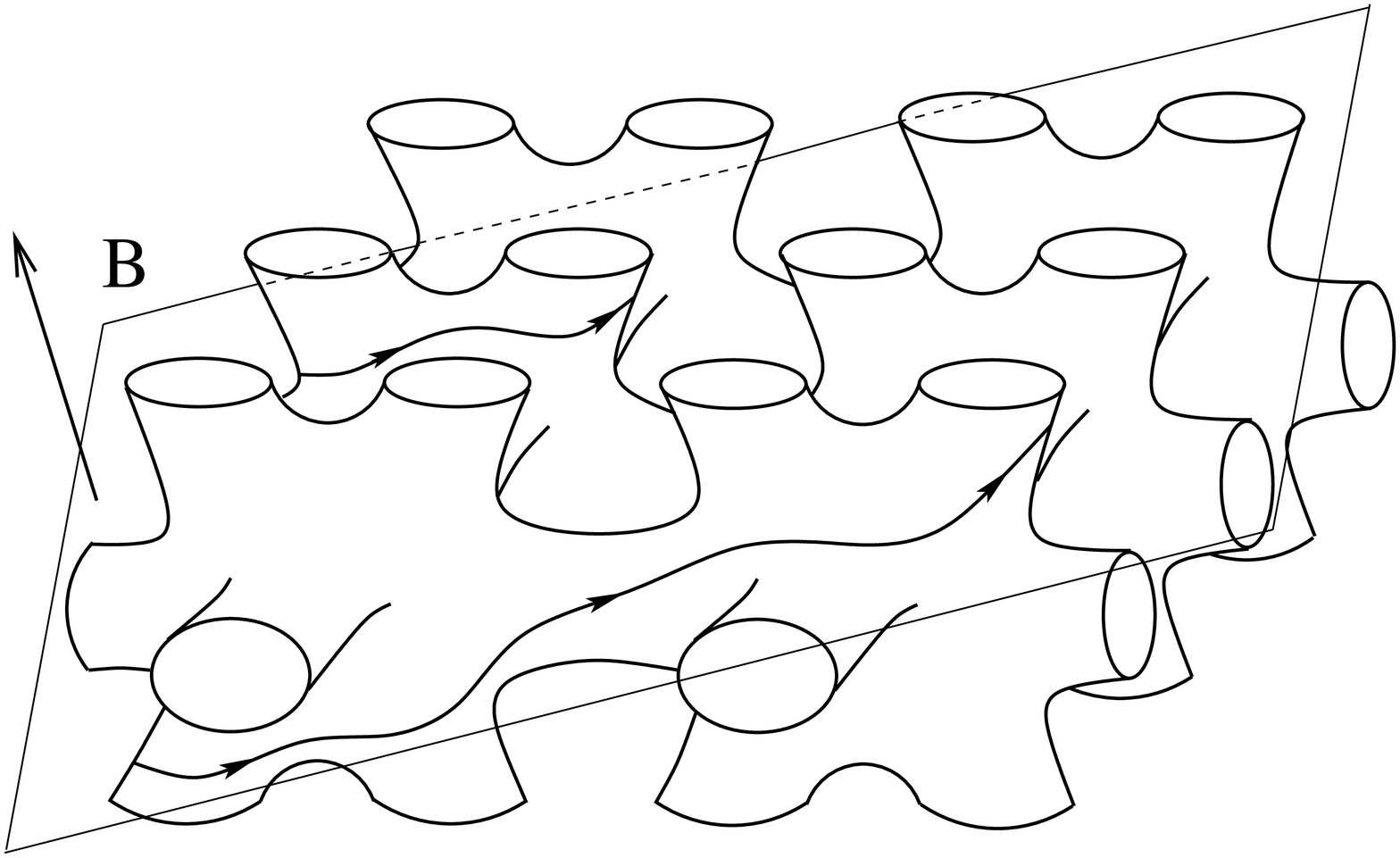}
\end{center}
\caption{The geometry of the trajectories of system (\ref{MFSyst}) 
in the space of quasimomenta.}
\label{Fig1}
\end{figure}

  In this work, we would like to consider features of oscillatory 
phenomena observed in reconstructions of the topological structure 
of the system describing the semiclassical motion of electrons on 
the Fermi surface in the presence of an external magnetic field. 
As is well known, this system has the form 
\begin{equation}
\label{MFSyst}
{\dot {\bf p}} \,\,\,\, = \,\,\,\, {e \over c} \,\,
\left[ {\bf v}_{\rm gr} ({\bf p}) \, \times \, {\bf B} \right]
\,\,\,\, = \,\,\,\, {e \over c} \,\, \left[ \nabla \epsilon ({\bf p})
\, \times \, {\bf B} \right] \,\,\, ,
\end{equation}
where $\, \epsilon ({\bf p}) \, $ is the electronic dispersion 
relation in the crystal for a given conduction band. The relation
$\, \epsilon ({\bf p}) \, $ represents a smooth 3-periodic function 
in the $\, {\bf p}$ - space with periods equal to the reciprocal 
lattice vectors. As it is easy to see, the system (\ref{MFSyst}) 
conserves the energy value $\, \epsilon ({\bf p}) \, $ and the 
projection of the quasimomentum on the direction of the magnetic 
field, and, as a consequence, its trajectories are given by the 
intersections of periodic surfaces
$\, \epsilon ({\bf p}) \, = \, {\rm const} \, $ by planes,
orthogonal to $\, {\bf B} \, $ (Fig. \ref{Fig1}).

 From the physical point of view, points in $\, {\bf p} $ - space, 
which differ by reciprocal lattice vectors, represent the same 
physical state, so that the system (\ref{MFSyst}) can be considered, 
in fact, as a system on the three-dimensional torus 
$\, \mathbb{T}^{3} \, $ obtained from $\, \mathbb{R}^{3} \, $ 
by factorization over reciprocal lattice vectors. The periodic 
surfaces $\, \epsilon ({\bf p}) \, = \, {\rm const} \, $ after 
such factorization become also compact two-dimensional surfaces 
embedded in $\, \mathbb{T}^{3} \, $ (as a rule, in a topologically 
nontrivial way). As is well known, in the theory of normal metals, 
among all energy levels, the Fermi energy plays the most important 
role, and thus the most important is the structure of the 
trajectories of system (\ref{MFSyst}) on the Fermi surface 
$\, \epsilon ({\bf p}) \, = \, \epsilon_{F} \, $.

 The great importance of the geometry of trajectories of system 
(\ref{MFSyst}) for the theory of galvanomagnetic phenomena in metals 
was established in the works of the school of I.M. Lifshits 
in the 1950s (see
\cite{lifazkag,lifpes1,lifpes2,lifkag1,lifkag2,lifkag3,etm,
KaganovPeschansky}). At this time many important and interesting 
examples of nontrivial behavior of the trajectories of system 
(\ref{MFSyst}) on complex Fermi surfaces, as well as the corresponding 
regimes of behavior of magnetic conductivity in strong magnetic fields, 
were considered. In the general case, the geometry of the trajectories 
of system (\ref{MFSyst}) starts to play a decisive role under the condition
$\, \omega_{B} \tau \gg 1 \, $, which also implies sufficient purity 
of the sample under study, as well as its low temperature
($T \sim 1 {\rm K}$) during the corresponding measurements.

  Somewhat later, in the work of S.P. Novikov 
\cite{MultValAnMorseTheory} the problem of general classification 
of trajectories of the system (\ref{MFSyst}) for arbitrary 
relations $\, \epsilon ({\bf p}) \, $ was set, which was then 
fruitfully investigated in his topological school 
(see \cite{zorich1,dynn1992,dynn1,zorich2,DynnBuDA,dynn2,dynn3}).
Topological results obtained at the school of S.P. Novikov, made 
it possible, in particular, to determine new topological 
characteristics observable in the conductivity of normal metals 
(\cite{PismaZhETF,UFN}), and also led to the discovery of new, 
previously unknown, types of trajectories of the system 
(\ref{MFSyst}) (\cite{Tsarev,dynn2}), leading to new regimes 
of behavior of magnetic conductivity (\cite{ZhETF1997,TrMian}).
On the whole, to date, it can be stated that the study of the 
Novikov problem has led ultimately to a complete classification 
of all types of trajectories of the system (\ref{MFSyst}), 
as well as to a description of the corresponding regimes of 
behavior of magnetic conductivity in strong magnetic fields 
(see, for example \cite{dynn3,UFN,BullBrazMathSoc,JournStatPhys,
UMNObzor,ObzorJetp}).

 It should be noted that a very important role in the study of 
the Novikov problem is played by the study of the set of closed 
trajectories of the system (\ref{MFSyst}) on the Fermi surface. 
Moreover, it can even be argued that knowledge of the structure 
of the set of closed trajectories on a given Fermi surface 
actually determines the types of all other trajectories on it and, 
in particular, allows one to describe their global geometric 
properties. It can also be noted that the set of nonsingular 
closed trajectories is always an open set on the Fermi surface 
and is locally stable with respect to small changes in the 
parameters of the problem (in particular, small changes in 
the Fermi energy or the direction of the magnetic field). 
From the above fact it follows, actually, that usually the 
space of parameters defining the system (\ref{MFSyst}) 
should be divided into regions in which the topological 
structure of system (\ref{MFSyst}) can be considered unchanged, 
while at the boundaries of such regions abrupt changes in 
the structure of (\ref{MFSyst}) occur. Besides that, any change 
in the structure of trajectories of (\ref{MFSyst}) on the Fermi 
surface is always associated with a reconstruction of the 
structure of closed trajectories on it, which, in fact, also 
determines the structure of other trajectories. 

\begin{figure}[t]
\begin{center}
\includegraphics[width=0.9\linewidth]{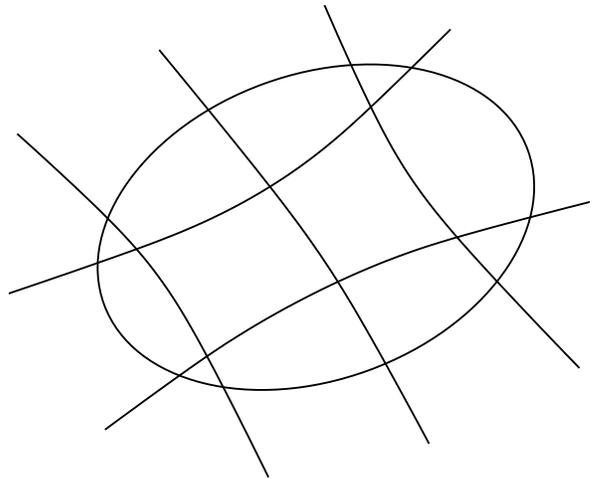}
\end{center}
\caption{Regions on a part of the unit sphere corresponding to 
different topological structures of system (\ref{MFSyst}) 
on the Fermi surface and the boundaries of reconstructions 
of this structure separating these regions (schematically).}
\label{Fig2}
\end{figure}

 In this work, we will be primarily interested in
the dependency of the topological structure of system
(\ref{MFSyst}) on the direction of the magnetic field 
(Fig. \ref{Fig2}). A typical picture of the boundaries 
separating different topological structures of 
(\ref{MFSyst}) on the corresponding angular diagram 
(on the unit sphere $\, \mathbb{S}^{2}$) was discussed in 
the most general case in \cite{OsobCycle}, where it was also 
indicated that the most convenient tool for observing it is 
the study of oscillatory phenomena (classical or quantum) 
for different directions of $\, {\bf B} $. The latter 
circumstance is due to the fact that changes in the 
topological structure of the system (\ref{MFSyst}) always 
cause the disappearance (and the appearance of new ones) 
of extreme trajectories, which play a central role in 
describing oscillatory phenomena in strong magnetic fields 
(cyclotron resonance, the de Haas - Van Alphen effect, 
the Shubnikov - de Haas effect, etc.). Thus, the boundaries 
separating different topological structures of system 
(\ref{MFSyst}) are, in fact, also the boundaries at which 
abrupt changes in the picture of classical or quantum 
oscillations occur under small changes of the direction 
of $\, {\bf B} $.

\begin{figure}[t]
\begin{center}
\includegraphics[width=0.9\linewidth]{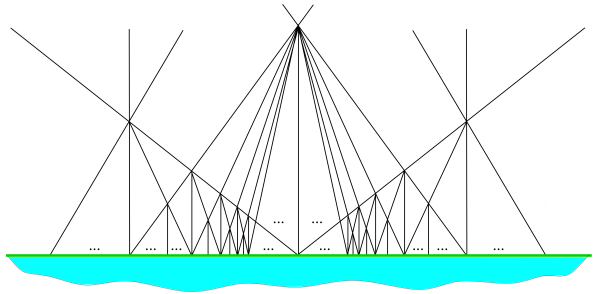}
\end{center}
\caption{Dense net of boundaries of elementary 
reconstructions of the structure of (\ref{MFSyst}), 
accumulating near a boundary of the appearance of 
open trajectories on the Fermi surface (schematically).}
\label{Fig3}
\end{figure}

 As was shown in \cite{OsobCycle}, the ``net'' of boundaries 
dividing the angular diagram into regions of a fixed topological 
structure of system (\ref{MFSyst}) is generally quite complex 
and consists of ``elementary'' segments, each of which corresponds 
to some ``elementary'' reconstruction of the structure of system  
(\ref{MFSyst}). The number of ``elementary'' segments can be 
generally infinite, in particular, the density of such segments 
becomes infinite near the directions of $\, {\bf B} \, $ 
corresponding to the appearance of open trajectories 
on the Fermi surface (Fig. \ref{Fig3}). The paper 
\cite{OsobCycle} also describes all the ``elementary'' 
reconstructions of the topological structure of (\ref{MFSyst}) 
on the Fermi surface that arise in the generic situation. 
Each of these reconstructions corresponds, in particular, 
to the disappearance and appearance of extreme trajectories 
of a very special form, determined by its topological type. 
As we have already said, each of the segments of the 
reconstructions of the structure of (\ref{MFSyst}) 
(one-dimensional curves in Fig. \ref{Fig2} and \ref{Fig3}) 
corresponds to an elementary reconstruction of a certain 
topological type, while topological types of reconstructions 
corresponding to different segments, generally speaking, 
are different.

 The main purpose of this work is to consider the features 
of observable oscillatory phenomena at the time of a change 
in the topological structure of the system (\ref{MFSyst}) 
on the Fermi surface. As we will see, each of the ``elementary'' 
reconstructions of this structure has its own peculiarities 
in the behavior of oscillations, which, in particular, can be very 
useful in the experimental determination of the topological types 
of such reconstructions.

\section{Topological types of elementary reconstructions and 
the distinctive features of the picture of oscillatory phenomena 
for reconstructions of different types}
\setcounter{equation}{0}

\begin{figure}[t]
\begin{center}
\includegraphics[width=\linewidth]{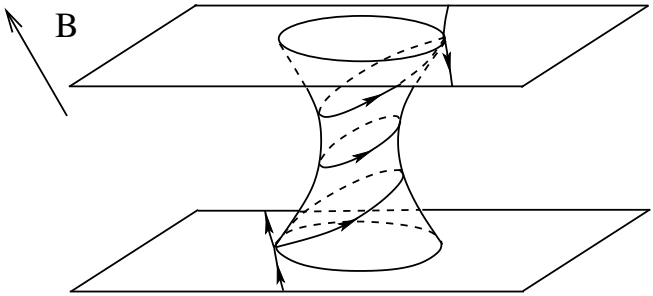}
\end{center}
\caption{A ``nontrivial'' cylinder of closed trajectories 
bounded by singular trajectories on its bases.}
\label{Fig4}
\end{figure}

  As we have already said, reconstructions of the topological 
structure of the system (\ref{MFSyst}) on the Fermi surface 
will mean for us topological reconstructions of the set of closed 
trajectories on this surface. In fact, as we have already noted 
above, the knowledge of the set of closed trajectories on the 
Fermi surface makes it possible to describe also trajectories 
of other types on it. The set of closed trajectories for generic 
directions of $\, {\bf B} \, $ is a finite set of (nonequivalent) 
cylinders bounded by singular closed trajectories on their bases 
(Fig. \ref{Fig4}). The structure of the set of cylinders of closed 
trajectories (their position on the Fermi surface and the scheme 
of their gluing with carriers of other trajectories and with each 
other) is locally stable for small rotations of $\, {\bf B} \, $ 
and can change only for special directions of $\, {\bf B} \, $ 
when it becomes a non-generic structure. More precisely, to change 
the topological structure of (\ref{MFSyst}), it is necessary to 
change the direction of $\, {\bf B} \, $ so that the height of 
one (or several) cylinder of closed trajectories becomes zero, i.e. 
we need the disappearance of a cylinder of closed trajectories, 
followed by the appearance of a new cylinder of low height 
(or several cylinders). The sets of directions of $\, {\bf B} \, $ 
corresponding to the moment of a reconstruction are 
one-dimensional curves on the angular diagram (on the unit sphere 
$\, \mathbb{S}^{2}$), whose union forms a ``net'' of directions of
$\, {\bf B} \, $, corresponding to the reconstructions of the 
structure of (\ref{MFSyst}) on the Fermi surface.

\begin{figure}[t]
\begin{center}
\includegraphics[width=\linewidth]{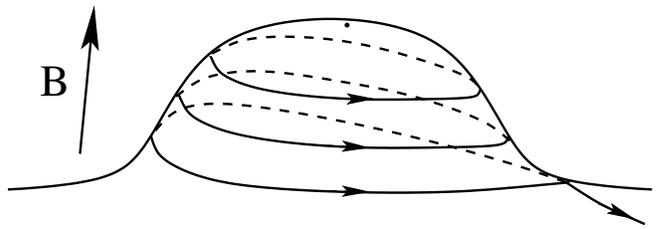}
\end{center}
\caption{A ``trivial'' cylinder of closed trajectories 
on the Fermi surface.}
\label{Fig5}
\end{figure}

\begin{figure}[t]
\begin{center}
\includegraphics[width=0.9\linewidth]{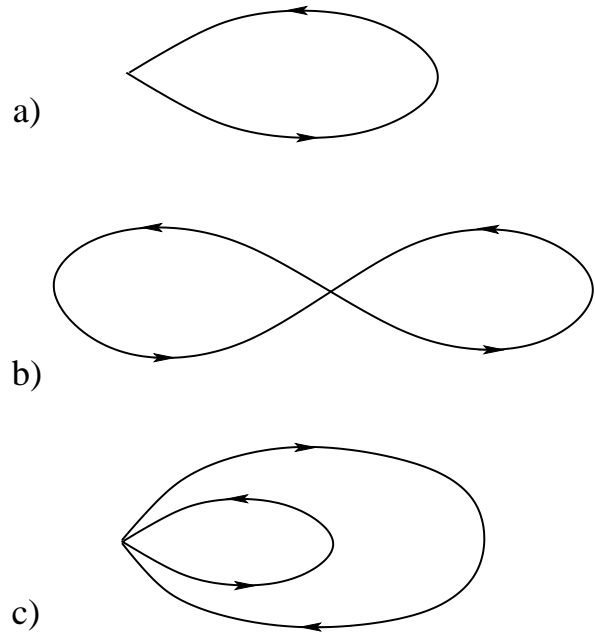}
\end{center}
\caption{Possible types of bases of ``nontrivial'' 
cylinders of closed trajectories. }
\label{Fig6}
\end{figure}

\begin{figure}[t]
\vspace{5mm}
\begin{tabular}{lc}
\includegraphics[width=0.5\linewidth]{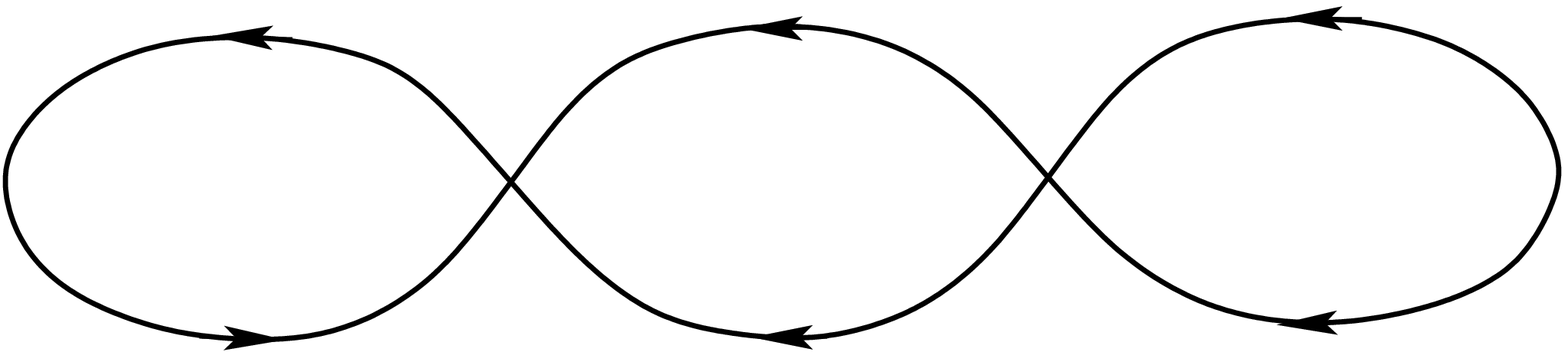}  &
\hspace{5mm}
\includegraphics[width=0.4\linewidth]{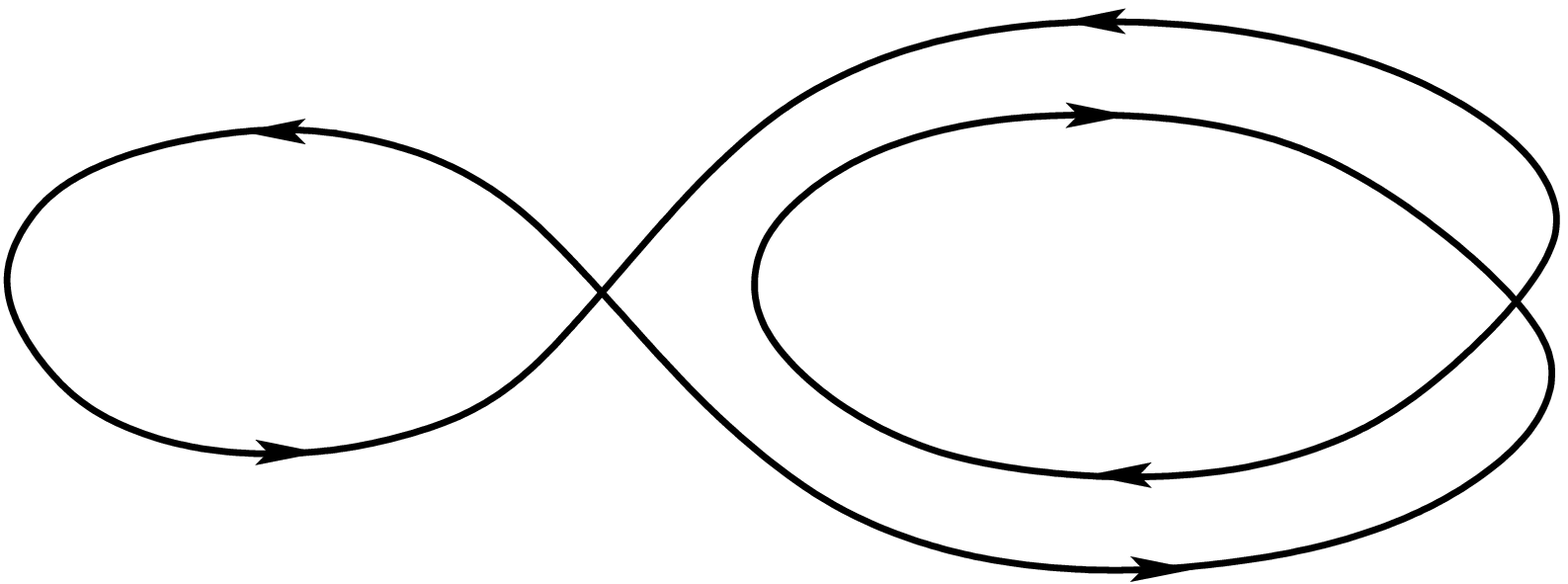}
\end{tabular}

\vspace{5mm}

\begin{tabular}{lc}
\includegraphics[width=0.4\linewidth]{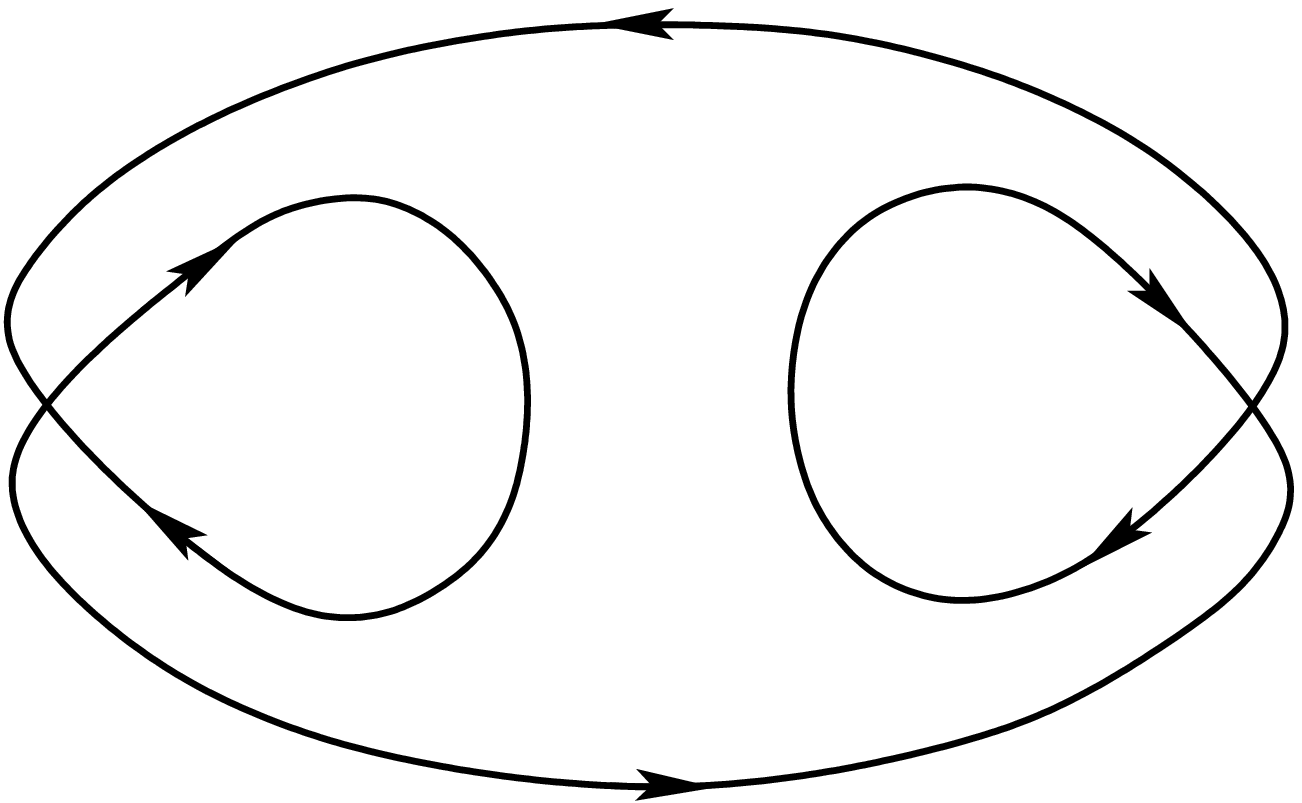}  &
\hspace{5mm}
\includegraphics[width=0.5\linewidth]{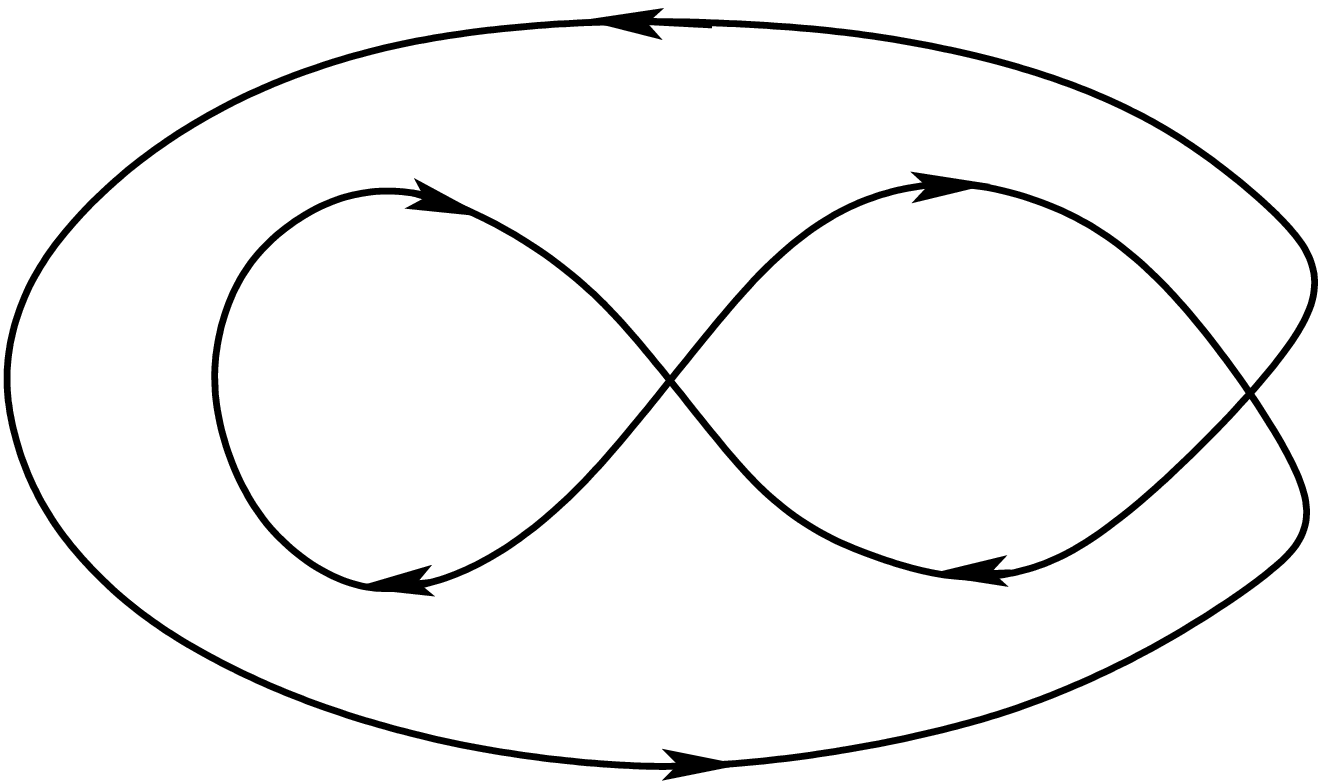}
\end{tabular}
\begin{tabular}{lc}
\includegraphics[width=0.5\linewidth]{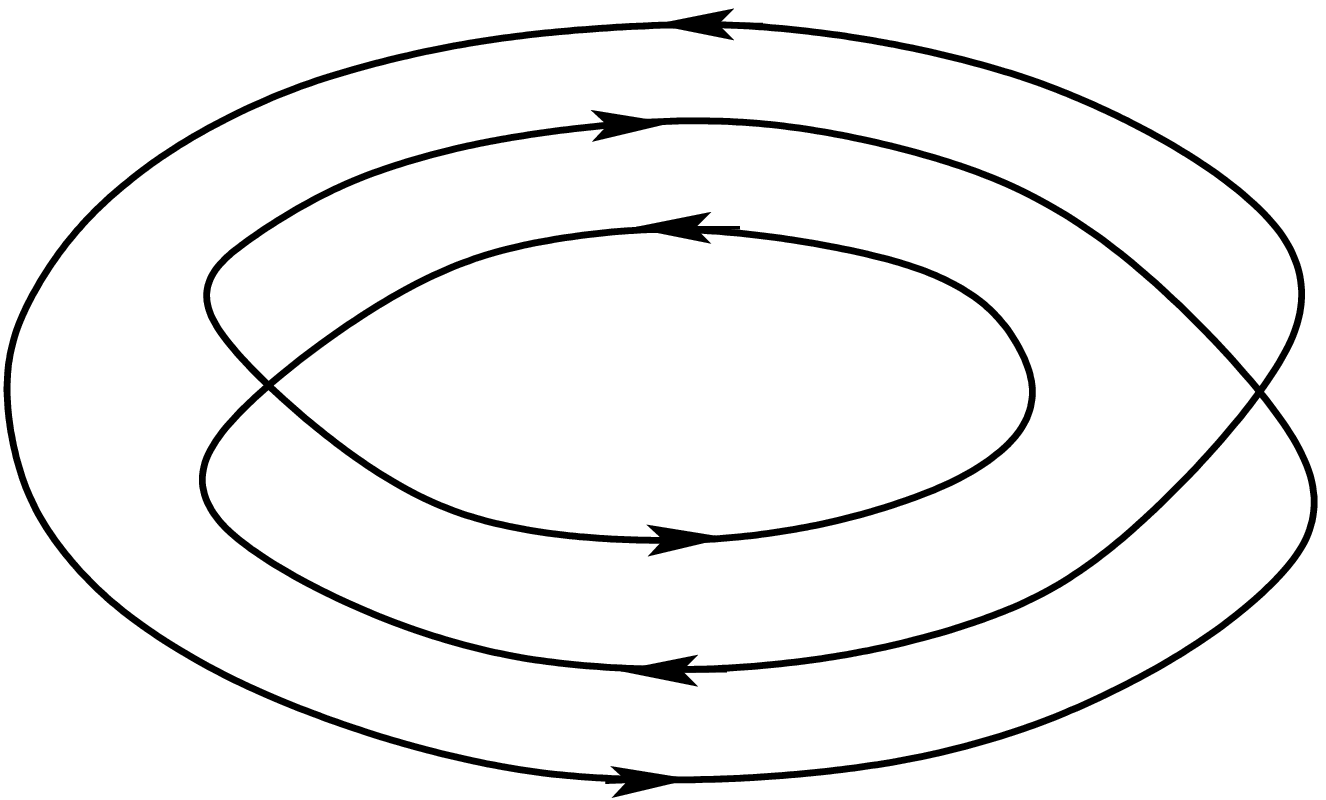}  &
\hspace{5mm}
\includegraphics[width=0.4\linewidth]{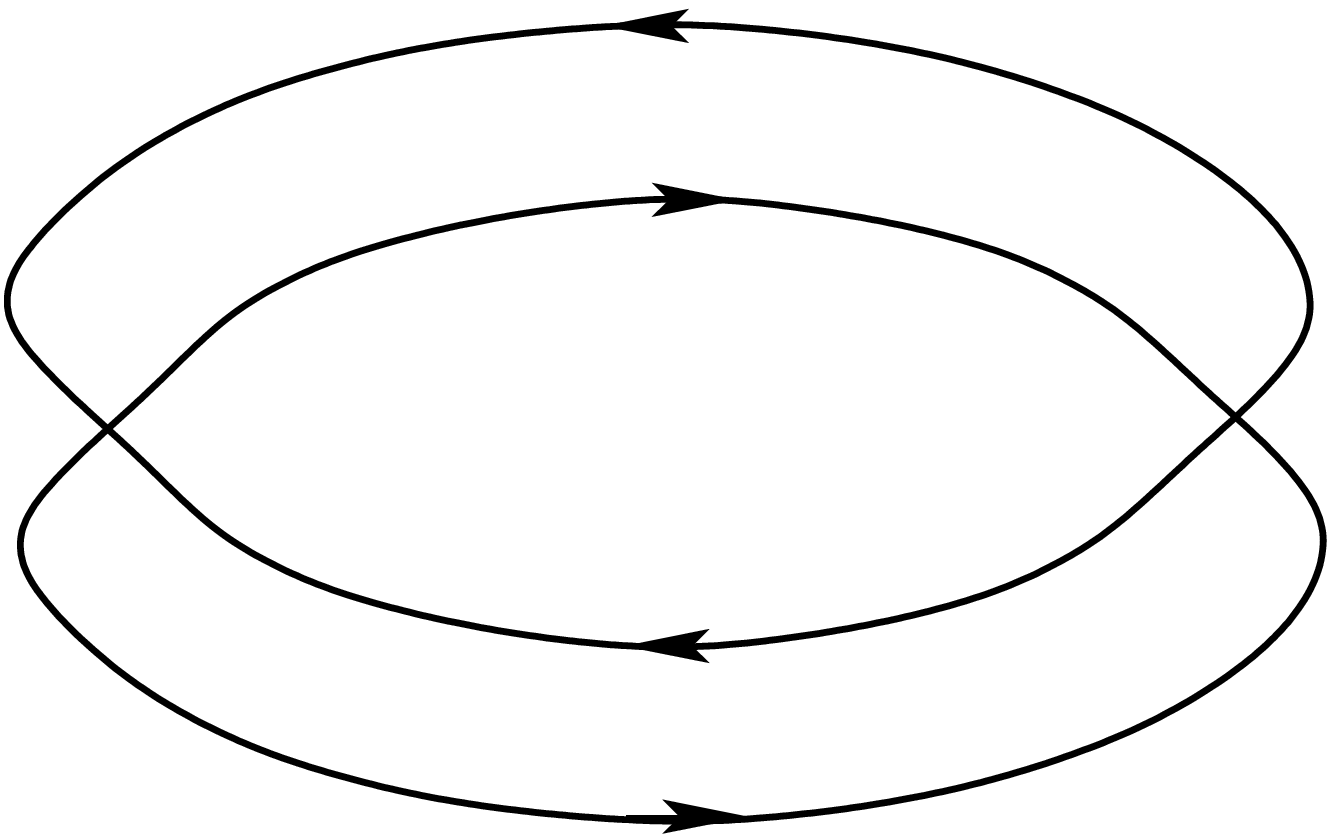}
\end{tabular}
\caption{``Cylinders of zero height'' arising at the 
moments of a reconstruction of the topological structure 
of system (\ref{MFSyst}) on the Fermi surface. }
\label{Fig7}
\end{figure}

 As in the work \cite{OsobCycle}, we will not pay 
attention here to the disappearance and appearance 
of ``trivial'' cylinders of closed trajectories, i.e. 
cylinders, at least one of the bases of which contracts 
to a single singular point (Fig. \ref{Fig5}), and we 
will consider only reconstructions of cylinders, both 
bases of which are ``nontrivial'' (Fig. \ref{Fig4}). 
In the generic case, we can assume that on each of the 
bases of such cylinders there is exactly one singular 
point of the system (\ref{MFSyst}), and each of the 
bases represents one of the figures shown at 
Fig. \ref{Fig6}. At the moment of reconstruction of 
the system (\ref{MFSyst}), a ``cylinder of zero height'', 
containing two singular points of (\ref{MFSyst}) 
connected by singular trajectories, appears on the Fermi 
surface. For each of the ``elementary'' reconstructions 
of the structure of (\ref{MFSyst}), the corresponding 
``zero-height cylinder'' represents a flat graph lying 
in a plane orthogonal to $\, {\bf B} \, $, and 
topologically is equivalent to one of the figures shown 
in Fig. \ref{Fig7}. As was shown in \cite{OsobCycle}, 
to determine the topological type of the ``elementary'' 
reconstruction of the system (\ref{MFSyst}), it suffices 
to fix the topological type of the corresponding 
``cylinder of zero height'' and indicate whether 
the group velocities at two singular points are 
co-directed, or directed opposite to each other.

\begin{figure}[t]
\begin{center}
\includegraphics[width=0.9\linewidth]{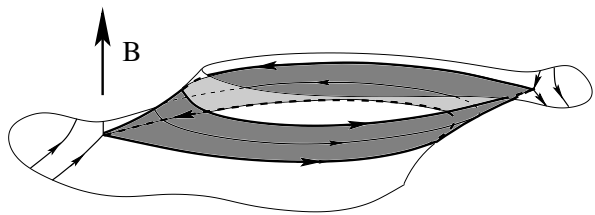}
\end{center}
\caption{Extremal closed trajectory on a vanishing (appearing) 
cylinder of closed trajectories near the moment of a reconstruction 
of the topological structure of system (\ref{MFSyst}) 
on the Fermi surface. }
\label{Fig8}
\end{figure}

 The most important circumstance in the situation under 
consideration is that on each of the cylinders of small height, 
up to its disappearance, there exist extreme closed trajectories 
of the system (\ref{MFSyst}) (having an extreme circulation 
period or area in comparison with close trajectories), 
disappearing together with the corresponding cylinder 
(Fig. \ref{Fig8}). When a new cylinder of closed trajectories 
appears, new extreme trajectories appear on it, which differ 
from the disappeared ones in their geometry. As a consequence, 
each reconstruction of the topological structure of system 
(\ref{MFSyst}) is followed by a sharp change in the picture 
of oscillatory phenomena in a strong magnetic field, which 
is a convenient tool for observing the above described 
``net'' of directions of $\, {\bf B} \, $ at the angular diagram. 
It must be said that extreme closed trajectories arising on 
cylinders of small height have certain peculiarities in 
comparison with ordinary extreme trajectories, namely, they 
contain sections that are very close to singular points of 
the system (\ref{MFSyst}). This circumstance leads, 
in particular, to an unlimited increase in the circulation 
period along such trajectories with a decrease in the height 
of the cylinder, as well as to a number of other features 
that arise, for example, when observing the phenomenon of 
cyclotron resonance (see, for example \cite{OsobCycle}).

 In this work, however, we would like to consider in more 
detail the features of extreme trajectories and the corresponding 
oscillatory phenomena that arise during each of the elementary 
reconstructions of the structure of (\ref{MFSyst}), which, 
from our point of view, can be rather useful in the experimental 
study of a complete picture of reconstructions of the topology 
of this system on complex Fermi surfaces. As is well known 
(see, for example, \cite{etm,Kittel,Abrikosov}), for description
of oscillatory phenomena, in fact, two types of closed extremal 
trajectories are important, namely, trajectories with an extreme 
orbital period and trajectories with an extreme area in comparison 
with trajectories close to them. Trajectories of the first type, 
as a rule, play a decisive role in the description of classical 
oscillatory phenomena (classical cyclotron resonance), while 
trajectories of the second type are important in describing 
quantum oscillatory phenomena (de Haas - van Alphen effect, 
Shubnikov - de Haas effect and etc.). Quite often, in reality, 
the same trajectory can be extreme from both the first and 
the second point of view, as a rule, this is the case for 
centrally symmetric sections of the Fermi surface. In most of 
the situations we consider below, however, this will not be 
the case, so we need to immediately divide the extreme 
trajectories into the two indicated types.

 As we have already said, we will consider here cylinders 
of closed trajectories with ``nontrivial'' bases containing 
one singular point of the system (\ref{MFSyst}). It is easy 
to see that the circulation period along closed trajectories 
on each of these cylinders increases infinitely (logarithmically) 
when approaching each of the bases. As a consequence, each 
of these cylinders must have at least one extremal trajectory 
with a minimum orbital period in comparison with trajectories 
close to it. 

\begin{figure}[t]
\begin{center}
\includegraphics[width=0.9\linewidth]{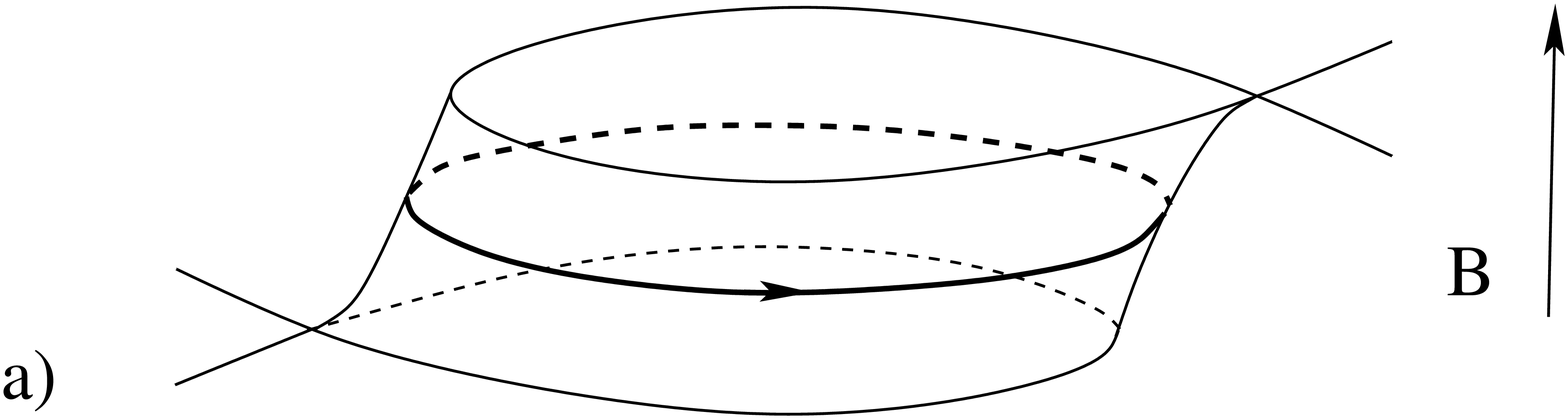}
\end{center}
\vspace{5mm}
\begin{center}
\includegraphics[width=0.9\linewidth]{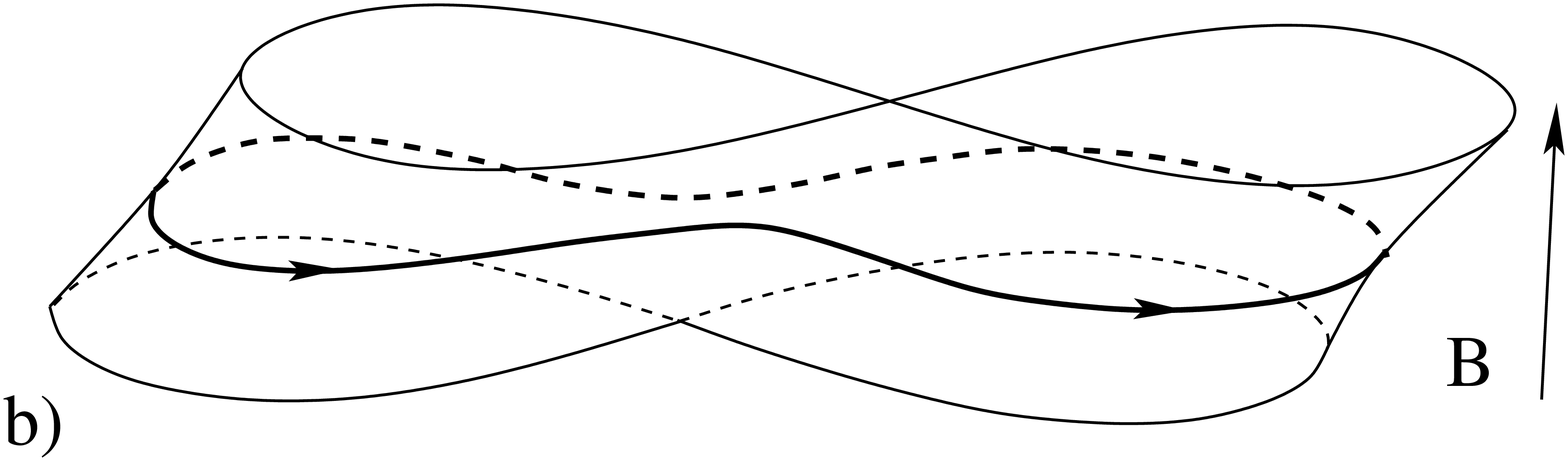}
\end{center}
\vspace{5mm}
\begin{center}
\includegraphics[width=0.9\linewidth]{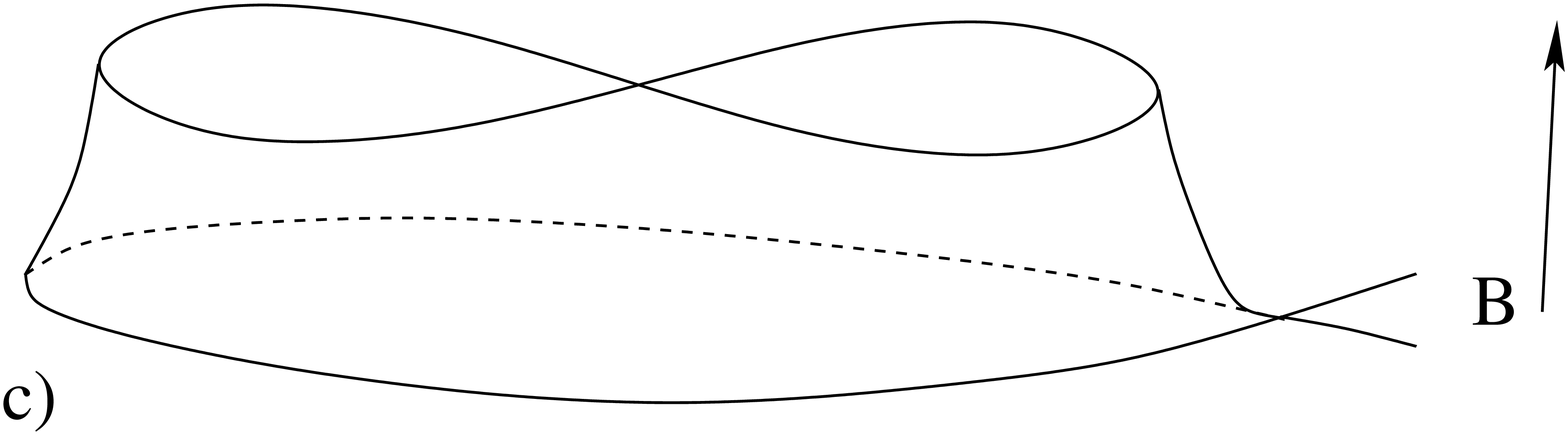}
\end{center}
\caption{Cylinders of closed trajectories containing 
trajectories of extreme (minimum and maximum) area (a and b), 
and a cylinder not containing trajectories of extreme area (c).}
\label{Fig9}
\end{figure}

 As for the area of closed trajectories, it is easy to see that 
it remains finite at the bases of the cylinders. Its derivative 
with respect to the distance to the corresponding base, however, 
goes to infinity (according to the logarithmic law) and can have 
a positive or negative sign, depending on the geometry of the 
cylinder. As for trajectories of the first type, this circumstance 
is also due to the presence of singular points on the bases of the 
cylinders and is associated with the local geometry of the 
trajectories near these points. Depending on the signs of the 
derivative of area with respect to the height on both bases, 
the cylinder of closed trajectories may or may not contain 
extreme trajectories of the second type. Fig. \ref{Fig9} shows 
examples of both cylinders containing extremal trajectories 
of the second type (a, b) and a cylinder that does not contain 
such a trajectory (c). It can be noted here that the extreme 
trajectory in Fig. \ref{Fig9}a  has a minimal area, while 
the extreme trajectory in Fig. \ref{Fig9}b  has the maximum 
area compared to trajectories close to them.

 Thus, it can be seen that any reconstruction of the structure 
of (\ref{MFSyst}) is always accompanied by a sharp change
in the picture of oscillations when observing, for example, 
the classical cyclotron resonance, while the picture of the
de Haas - van Alphen or the Shubnikov - de Haas oscillations 
may contain no abrupt changes (if cylinders of small height 
on both sides of the reconstruction do not contain trajectories 
of extreme area). Special note should be made about the 
reconstructions of (\ref{MFSyst}), which have central symmetry. 
In this case, the central sections of cylinders of small height 
always represent extreme trajectories of both the first and the 
second type. 

 In the most general case, cylinders of small height can contain 
extreme trajectories of both types, which, however, do not coincide 
with each other. In this case, although the reconstruction of 
(\ref{MFSyst}) is accompanied by sharp changes in the pictures 
of oscillations of all types, one can observe a difference in the 
parameters of the corresponding disappearing or new oscillatory 
terms. For example, when observing the phenomenon of cyclotron 
resonance, the orbital period is directly measured along extreme 
trajectories, which give the main terms in the overall picture of 
oscillations. At the same time, the orbital period can also be 
measured for trajectories of extreme area, for example, from the 
temperature dependence of the corresponding quantum oscillations 
(\cite{etm,LifshitzKosevich1954}). It is easy to see that these 
quantities must coincide in the case when both types of oscillations 
are generated by the same trajectory and differ if different types of 
oscillations correspond to different extremal trajectories.

\begin{figure}[t]
\begin{center}
\includegraphics[width=0.9\linewidth]{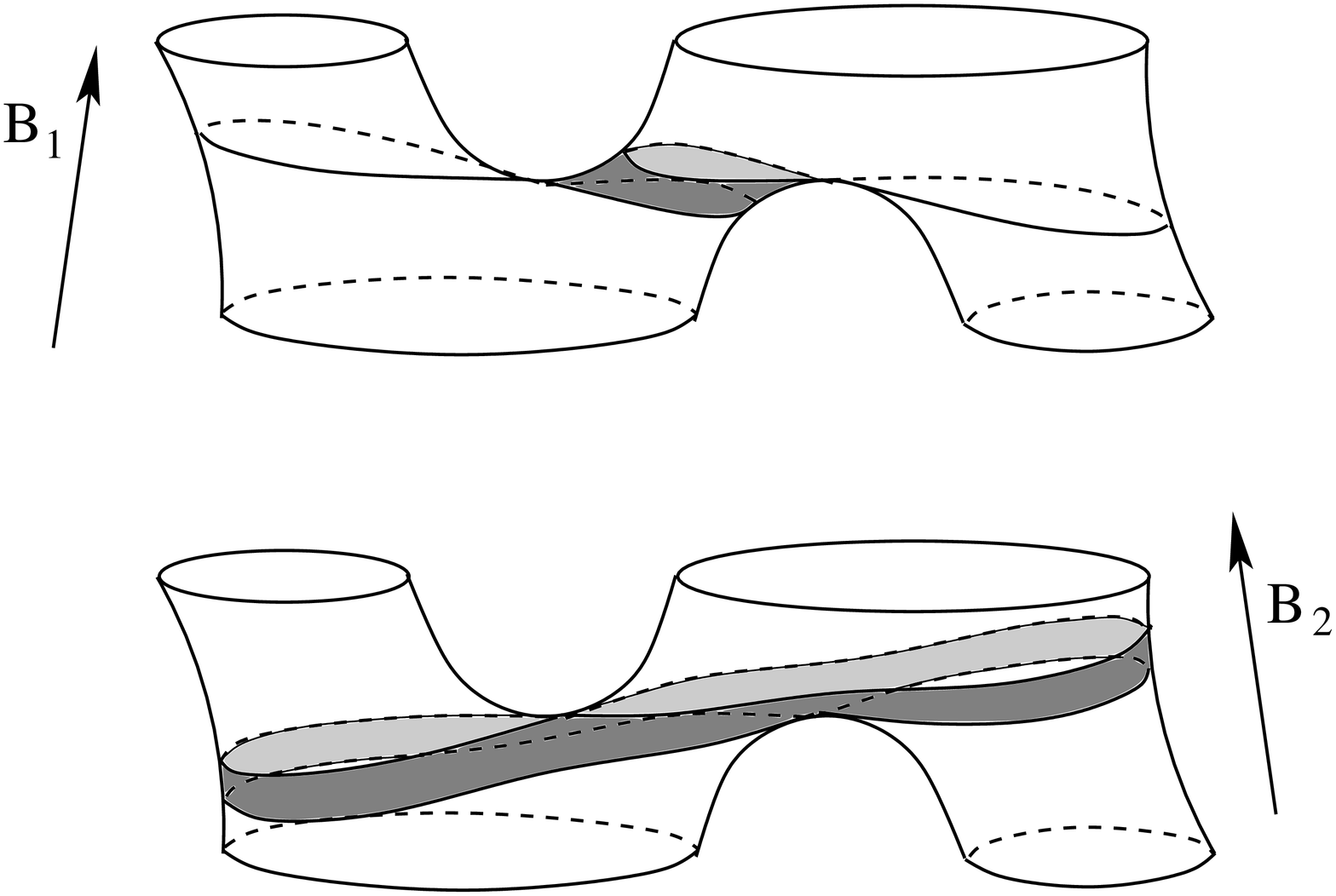}
\end{center}
\vspace{5mm}
\begin{center}
\includegraphics[width=0.8\linewidth]{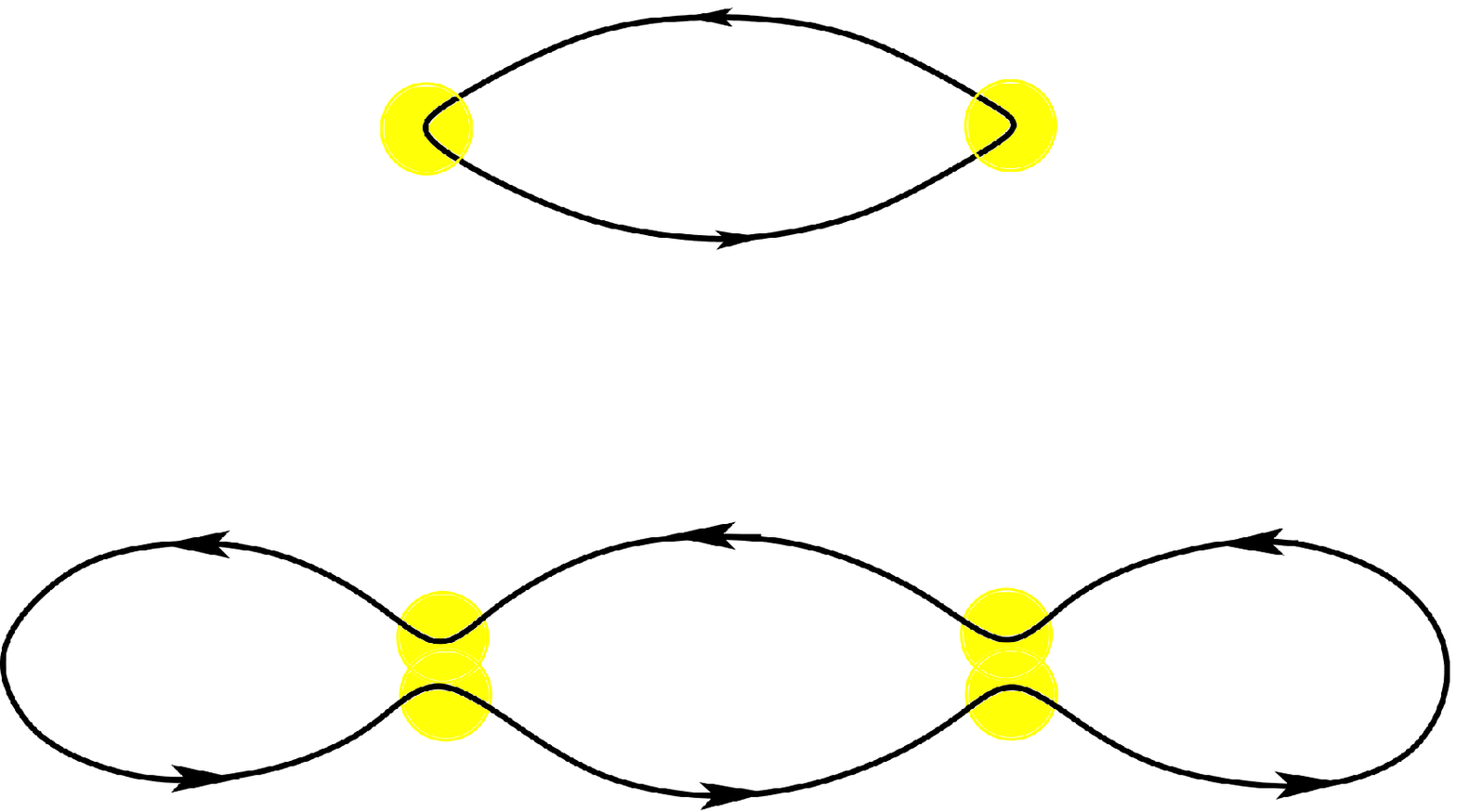}
\end{center}
\caption{One of the most common reconstructions of the topological 
structure of system (\ref{MFSyst}) on the Fermi surface and extreme 
trajectories on cylinders of small height, before and after 
the reconstruction (the color indicates the areas close to the 
singular points of system (\ref{MFSyst})). }
\label{Fig10}
\end{figure}

\begin{figure}[t]
\begin{center}
\includegraphics[width=0.9\linewidth]{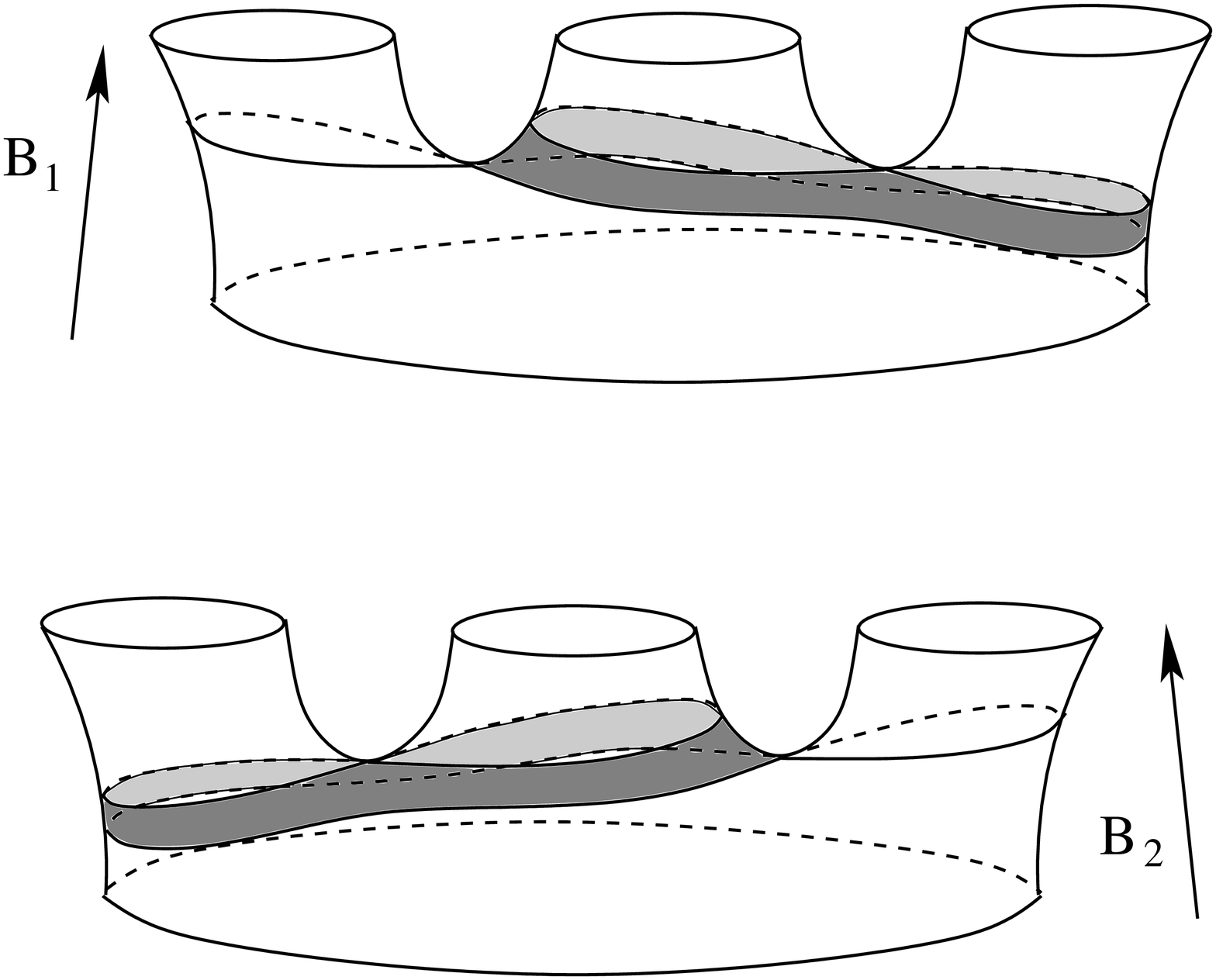}
\end{center}
\vspace{5mm}
\begin{center}
\includegraphics[width=0.9\linewidth]{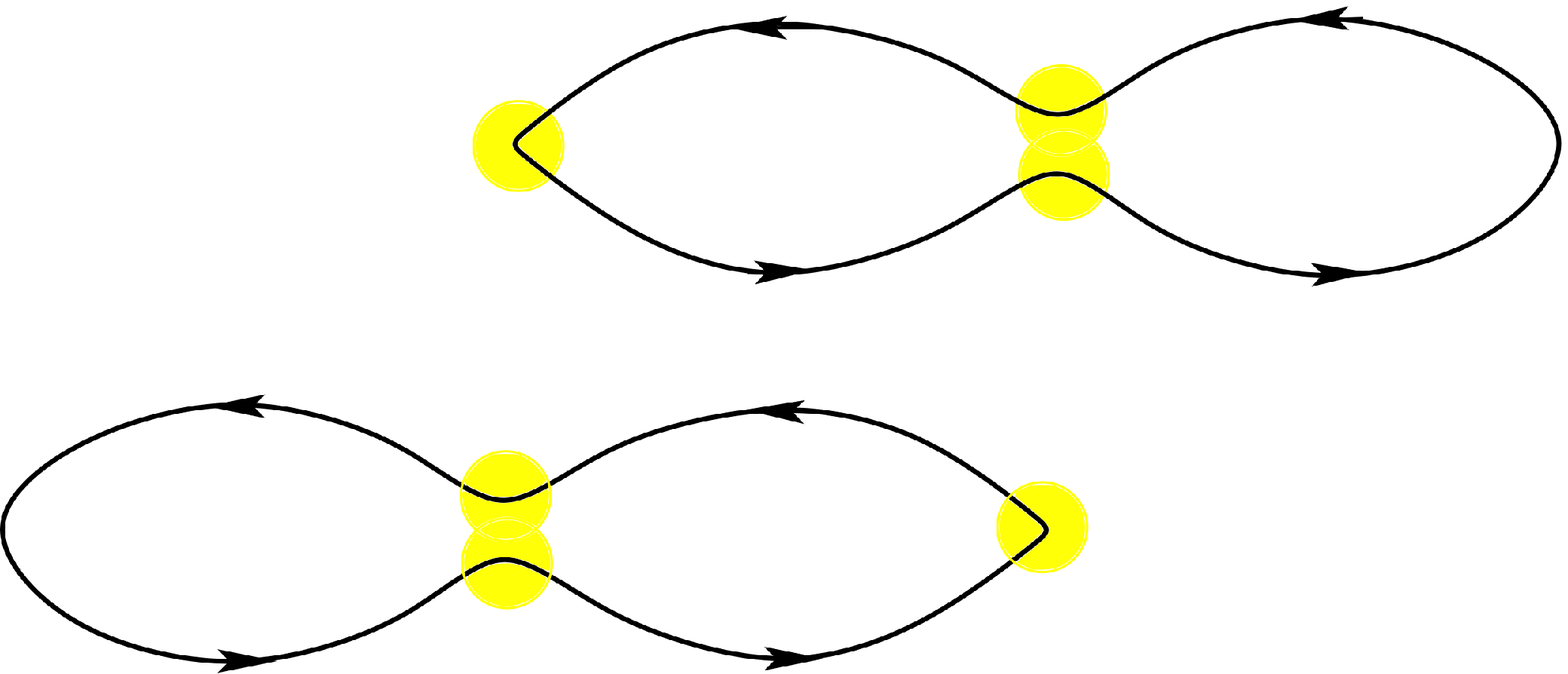}
\end{center}
\caption{One of the possible reconstructions of the topological 
structure of system (\ref{MFSyst}) on the Fermi surface and 
extremal trajectories on cylinders of low height, before and 
after the reconstruction (the sections close to the singular 
points of system (\ref{MFSyst}) are marked with color).}
\label{Fig11}
\end{figure}

 As an example, consider two different reconstructions 
shown in Fig. \ref{Fig10} and \ref{Fig11}. Both reconstructions 
actually correspond to the same topology of 
a ``cylinder of zero height'' (the first of those shown 
in Fig. \ref{Fig7}) and differ only in the directions of group 
velocities at two saddle singular points of the system 
(\ref{MFSyst}) (oppositely directed and co-directed velocities 
at singular points).

 The reconstruction shown in Fig. \ref{Fig10} may have central 
symmetry and, thus, it may appear on one part of the Fermi surface 
(the most common case). Though, the topological structure at 
Fig. \ref{Fig10} also may not have central symmetry. In this case, 
it should appear simultaneously on two parts of the Fermi surface, 
transforming into each other under the central inversion in the 
$\, {\bf p} $ - space. Whether or not the structure at 
Fig. \ref{Fig10} has central symmetry, on the corresponding cylinders 
of small height, both before and after the reconstruction, trajectories 
of extreme area appear, and one of them (before the reconstruction) 
has the minimum area, while the second one (after the reconstruction) 
has the maximum area in comparison with trajectories close to them. 
Thus, the reconstruction shown in Fig. \ref{Fig10} should always be 
accompanied by both a sharp jump in one of the oscillating terms in 
classical oscillations (change in the geometry of a trajectory 
of the extreme period), and a sharp jump in one of the oscillating 
terms in quantum oscillations (change in the geometry of a trajectory 
of the extreme area). Both the trajectories of the extreme period 
and the trajectories of the extreme area have here the shape shown 
in Fig. \ref{Fig10}, while in the case of the presence of central 
symmetry, they simply coincide. As we said above, in the latter case, 
the orbital periods measured from the classical oscillations and the 
temperature dependence of the quantum oscillations of the corresponding 
oscillatory terms must coincide.

 The reconstruction shown in Fig. \ref{Fig11}, cannot have central 
symmetry and its appearance is possible only in pairs, on the 
sections of the Fermi surface that transform into each other under the 
central inversion in the $\, {\bf p} $ - space. The cylinders of low height, 
both before and after the reconstruction, coincide with the one shown 
at the last figure of Fig. \ref{Fig9}, and do not contain extreme area 
trajectories. On these cylinders, however, there are always trajectories 
with an extreme orbital period, the shape of which is shown at 
Fig. \ref{Fig11}. When crossing the boundary of such a reconstruction, 
therefore, a jump occurs (a sharp change in one of the oscillatory terms) 
only in the picture of classical oscillations (classical cyclotron 
resonance, etc.).

 Below, Fig. \ref{Fig12} - \ref{Fig21} present all the 
remaining topological types of ``elementary'' reconstructions 
of system (\ref{MFSyst}). In addition to the shape of a 
reconstruction itself, Fig. \ref{Fig12} - \ref{Fig21} also show 
the shape of the lower and upper bases of cylinders of small 
height before and after the resconstruction ((a, b) and (d, e)), 
the shape of extreme trajectories on cylinders of small height 
((c) and (f)), as well as the structure of the 
``cylinder of zero height'' that appears immediately at the 
moment of the reconstruction (g). Strictly speaking, the above 
figures accurately depict the local geometry of the extreme 
trajectories near the above-mentioned ``deceleration sections'' 
on them (colored sections), as well as the topology of their 
connection by the remaining trajectory sections, but in other 
details, they can be geometrically more complicated. For the 
cylinders of small height under consideration, the extremal 
trajectories of both types are geometrically very close to each 
other in $\, {\bf p} $ - space (if both types of trajectories 
are present on the cylinder), but they can differ significantly 
from each other in other parameters (for example, the value 
of the circulation period along the trajectory). As we said 
above, the main goal of this work is to describe the features 
of oscillatory (and other) phenomena that make it possible to 
distinguish between different types of ``elementary reconstructions'' 
of the system (\ref{MFSyst}) during their experimental observation.

 Fig. \ref{Fig12} - \ref{Fig16} represent the reconstructions, 
during which trajectories of extreme area do not appear and 
do not disappear on the corresponding cylinders of small height. 
The extreme trajectories shown at these figures have just 
the smallest orbital period among all cylinder trajectories. 
Together with the reconstruction shown in Fig. \ref{Fig11}, 
such reconstructions can be attributed 
to ``reconstructions of the first group''. As we said above, 
reconstructions of this type are distinguished by the fact that 
during them we have a sharp change of some of the oscillatory 
terms only in the picture of classical oscillations.

 As it is easy to check, in all the reconstructions shown 
in Fig. \ref{Fig11} - \ref{Fig16}, group velocities at two 
saddle singular points are co-directed to each other. 
Coming back to the description of elementary reconstructions 
in terms of the topology of ``cylinders of zero height'' 
(Fig. \ref{Fig7}), it is easy to formulate a simple rule. 
Namely, for any of the types of ``zero-height cylinders'' 
shown in Fig. \ref{Fig7}, trajectories of extreme area 
appear on the corresponding cylinders of small height 
(both before and after the reconstruction) if the group 
velocities at its singular points are directed opposite 
to each other.

 The above rule can be easily justified using the 
well-known formula 
$${\partial S \over \partial p_{z}} \,\,\, = \,\,\,
{e B \over c} \, \oint v_{gr}^{z} \, d t $$
(where $\, t \, $ is the time of motion along the trajectory) 
for the area of a closed trajectory in $\, {\bf p}$ - space
$\, S (p_{z}) \, $. Since the singular points near the 
reconstruction are located at the bases of cylinders of small 
height, and the time of their passage tends to infinity when 
approaching the bases of a cylinder, this relation determines 
the signs of $\, \partial S / \partial p_{z} \, $ 
near the bases of the cylinders. As is also well known, 
on trajectories of the extreme area, we have the relation 
$$ \oint v_{gr}^{z} \, d t \,\,\, = \,\,\, 0 $$

\begin{figure}[t]
\begin{center}
\includegraphics[width=0.9\linewidth]{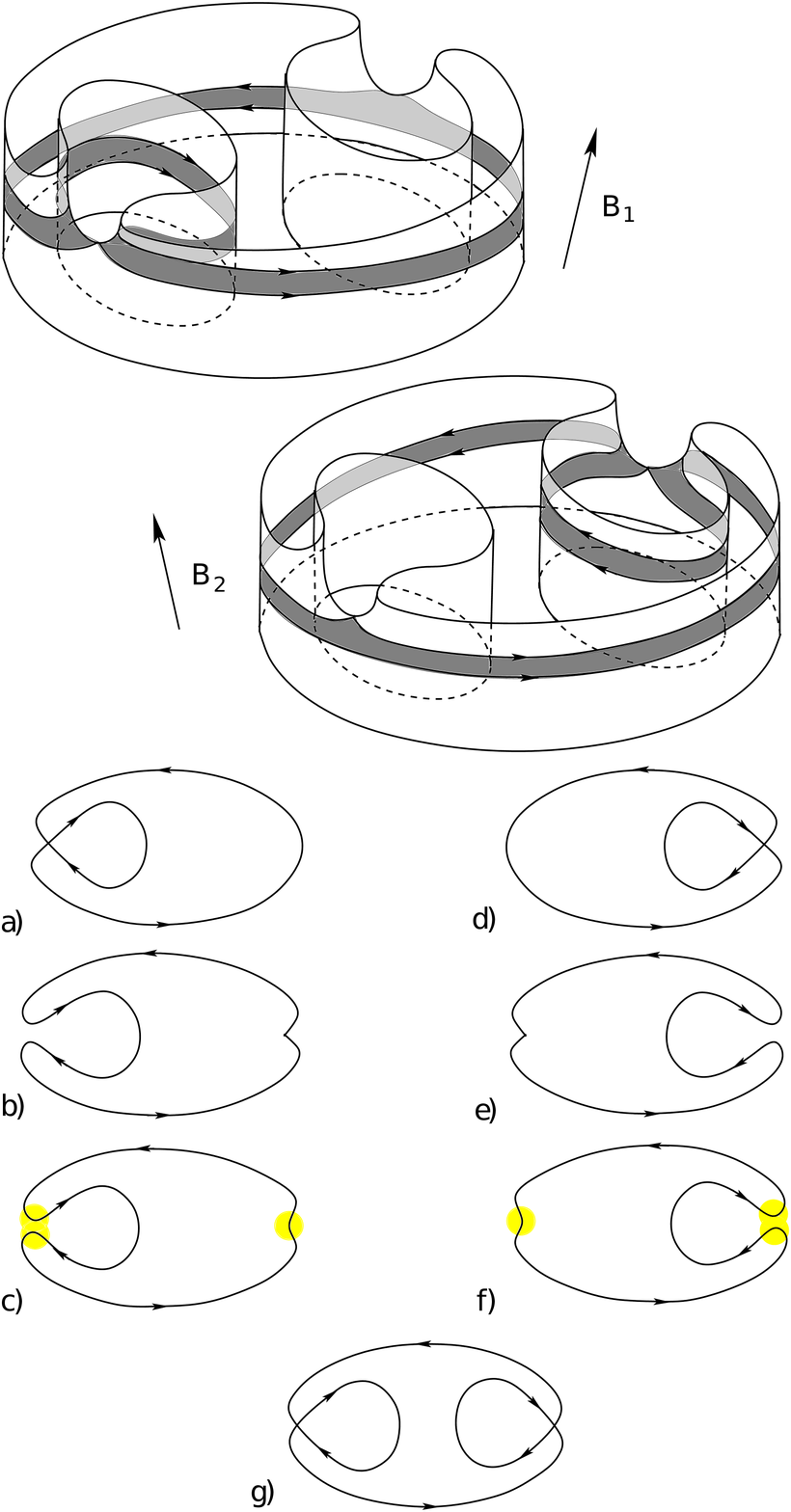}
\end{center}
\caption{Reconstruction of the structure of (\ref{MFSyst}) 
without central symmetry. The trajectories of the extreme 
area on cylinders of low height are absent both before and 
after the reconstruction. The trajectories of the minimum 
orbital period on opposite sides of the reconstruction 
are of the same type (electron or hole). }
\label{Fig12}
\end{figure}

\begin{figure}[t]
\begin{center}
\includegraphics[width=0.9\linewidth]{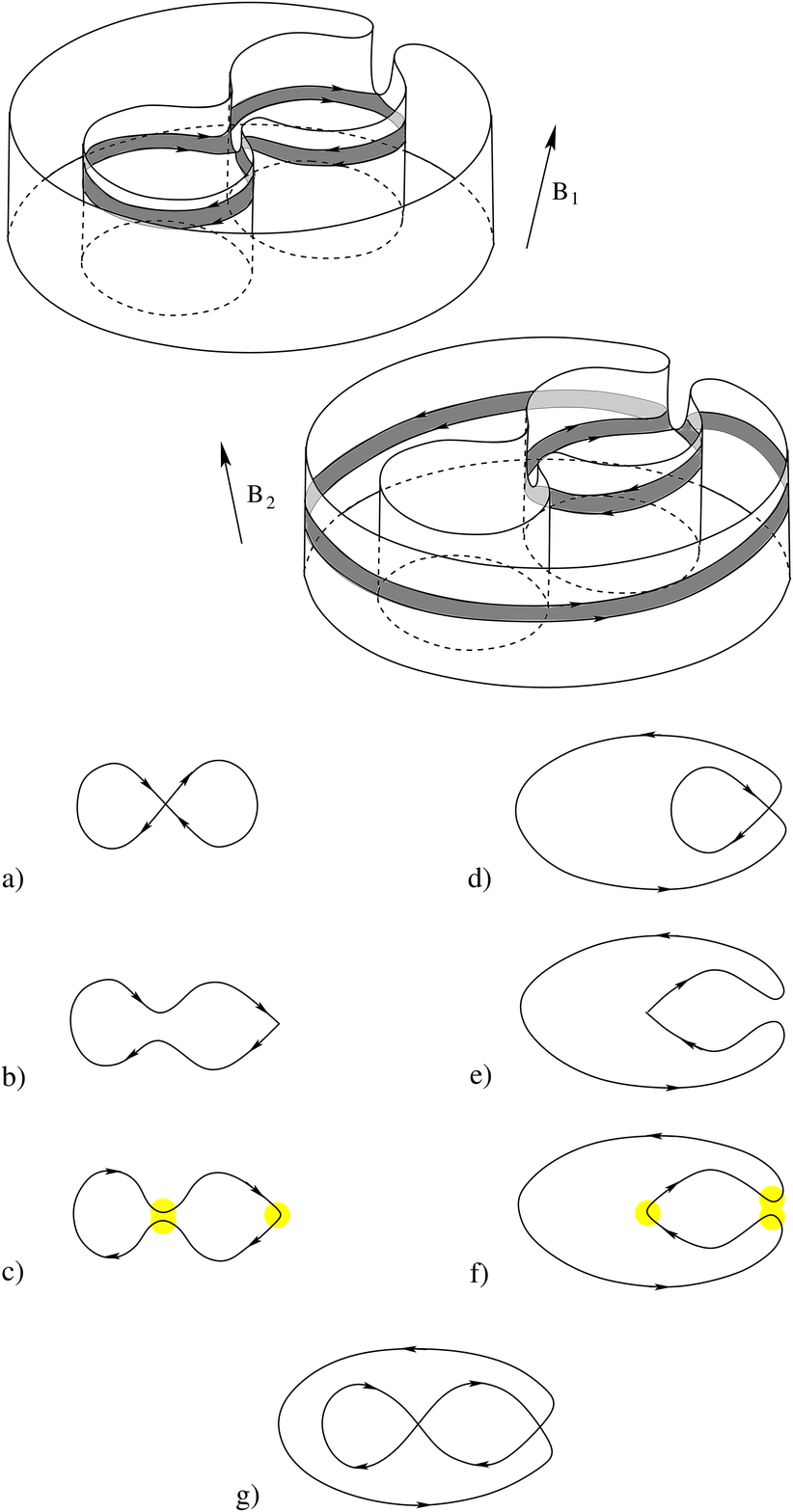}
\end{center}
\caption{Reconstruction of the structure of (\ref{MFSyst}) 
without central symmetry. The trajectories of the extreme 
area on cylinders of low height are absent both before and 
after the reconstruction. The trajectories of the minimum 
orbital period on opposite sides of the reconstruction are 
of different types (electronic, on the one side of the 
reconstruction, and hole, on the other). }
\label{Fig13}
\end{figure}

 It can be noted that the extreme trajectories shown at 
Fig. \ref{Fig12} - \ref{Fig13}, as well as the trajectories 
shown in Fig. \ref{Fig11}, approach the saddle singular points 
of the system (\ref{MFSyst}) three times (twice to one of the 
singularities and once to the second). Besides that, on different 
sides of the reconstruction, the multiplicity of the approach 
of a special trajectory to each of the singular points changes 
(before the reconstruction, the trajectory approaches twice 
to one of the singular points, and after the reconstruction, 
to the other). Each approach to a singular point of 
(\ref{MFSyst}) means the presence of a ``deceleration section'' 
on the corresponding part of a trajectory, which gives a finite 
addition to the trajectory orbital period. The corresponding 
addition to the orbital period grows logarithmically with 
decreasing angle $\, \alpha \, $  between the direction of
$\, {\bf B} \, $ and the boundary of a reconstruction 
and for $\, \alpha \ll 1 \, $ can be written as
$$\Delta T_{i} \,\,\, \simeq \,\,\, 
{c \over e B v^{(i)}_{gr}} \, {1 \over \sqrt{G_{i}}} \,\,
\ln {1 \over \alpha} \,\,\, , $$
where $\, v^{(i)}_{gr} \, $ and $\, G_{i} \, $ are, 
respectively, the values of the group velocity and the 
Gaussian curvature of the Fermi surface at each of the 
singular points ($i = 1,2$). 

 The total values of $\, \Delta T (\alpha) \, $ can be 
measured with a sufficiently accurate measurement of the 
orbital period in classical oscillatory phenomena and 
a sufficiently close approximation of the direction of 
$\, {\bf B} \, $ to the boundary of a reconstruction of 
the structure of system (\ref{MFSyst}). It is easy to see 
then that when crossing the boundary, the corresponding total 
addition to the period of circulation along an extreme 
trajectory changes from the value
$\, \Delta T (\alpha) = 2 T_{1} (\alpha) + T_{2} (\alpha) \, $
to the value $\, T_{1} (\alpha) + 2 T_{2} (\alpha) \, $,
which distinguishes the reconstructions at 
Fig. \ref{Fig11} - \ref{Fig13} among all the reconstructions
shown in Fig. \ref{Fig11} - \ref{Fig16}. To distinguish 
different reconstructions in Fig. \ref{Fig11} - \ref{Fig13} 
a test can be used which determines the possibility of
deceleration sections falling into the skin layer on each 
side of the boundary of a reconstruction 
(see \cite{OsobCycle}).

\begin{figure}[t]
\begin{center}
\includegraphics[width=0.9\linewidth]{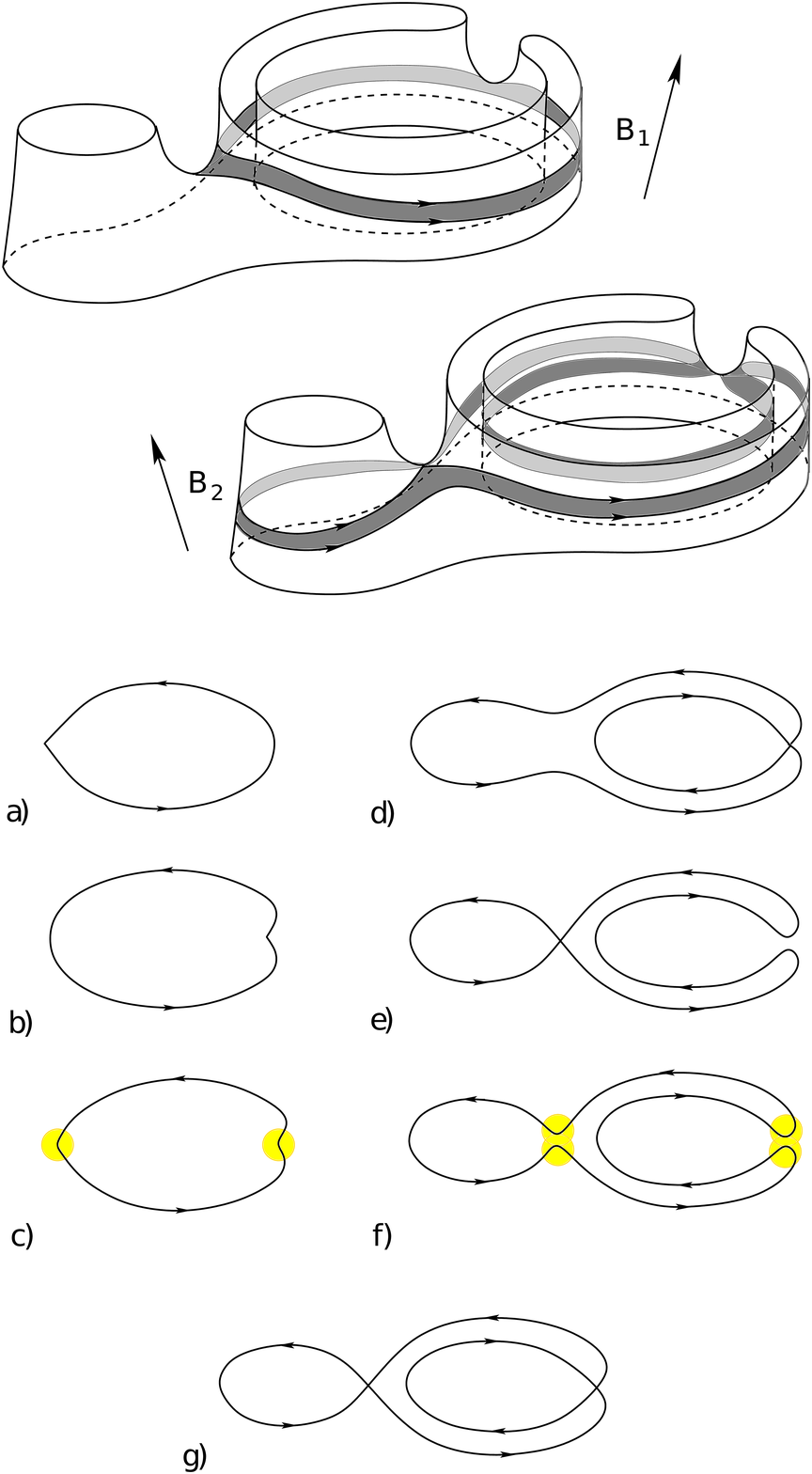}
\end{center}
\caption{Reconstruction of the structure of (\ref{MFSyst}) 
without central symmetry. The trajectories of the extreme 
area on cylinders of low height are absent both before and 
after the reconstruction. The trajectories of the minimum 
orbital period on opposite sides of the reconstruction are 
of the same type (electron or hole).}
\label{Fig14}
\end{figure}

\begin{figure}[t]
\begin{center}
\includegraphics[width=0.9\linewidth]{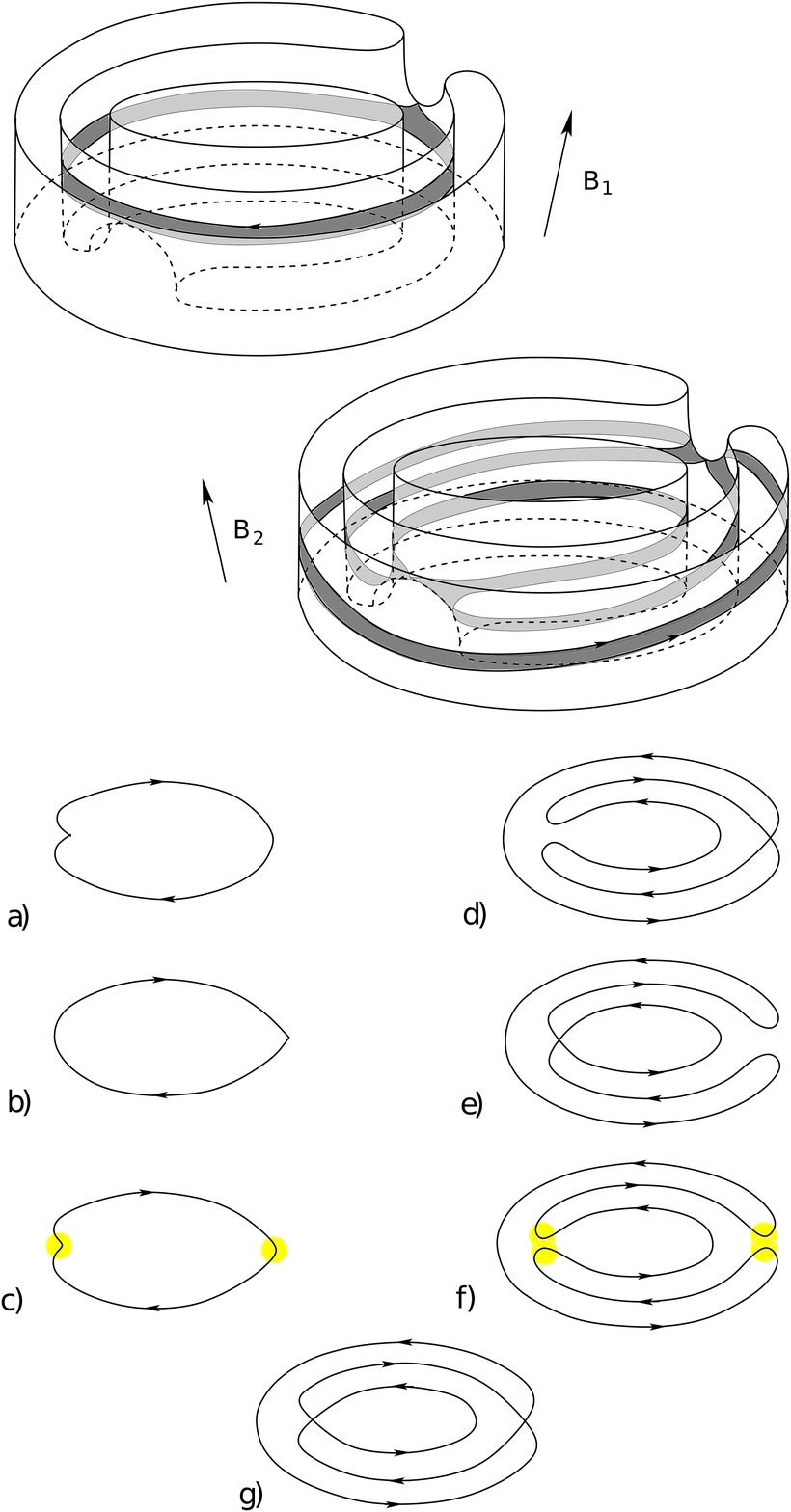}
\end{center}
\caption{Reconstruction of the structure of (\ref{MFSyst}) 
without central symmetry. The trajectories of the extreme 
area on cylinders of low height are absent both before and 
after the reconstruction. The trajectories of the minimum 
orbital period on opposite sides of the reconstruction are 
of different types (electronic, on the one side of the 
reconstruction, and hole, on the other).}
\label{Fig15}
\end{figure}

 For the reconstructions shown in Fig. \ref{Fig14} - \ref{Fig15}, 
we have a different situation. Namely, now extreme trajectories 
have two ``deceleration segments'' on one side of a reconstruction 
and four ``deceleration segments'' on the other. It is easy to see 
that when crossing the corresponding boundary of the reconstruction 
the total addition to the orbital period due to the 
``deceleration segments'' changes from the value 
$\, \Delta T (\alpha) = T_{1} (\alpha) + T_{2} (\alpha) \, $
to the value
$\, 2 \left( T_{1} (\alpha) + T_{2} (\alpha) \right) \, $,
which also makes it possible to distinguish these reconstructions
among the six reconstructions of the first group.

 To distinguish two reconstructions presented in 
Fig. \ref{Fig14} - \ref{Fig15} we can use, for example, 
their geometric (and topological) differences in the 
$\, {\bf p} $ - space, which are also transferred to 
the coordinate space. For example, for the trajectory 
shown in Fig. \ref{Fig14}c, most of its sections can be located 
in the skin layer at the sample boundary both before crossing 
the reconstruction boundary and immediately after crossing it 
(we assume that the magnetic field is directed parallel to the 
sample boundary). For the trajectory shown in Fig. \ref{Fig15}c, 
such sections do not exist, which is due to a significantly 
different topology of its reconstruction. The above difference 
for the considered reconstructions can be established, for example, 
by the absence or presence of a jump in the direction of 
$\, {\bf v}_{gr} \, $ in the section falling into the skin 
layer when crossing the boundary of a reconstruction. 
We also note here that the measurement of the corresponding 
direction of $\, {\bf v}_{gr} \, $ is almost always performed 
in the observation of the classical cyclotron resonance.

\begin{figure}[t]
\begin{center}
\includegraphics[width=0.9\linewidth]{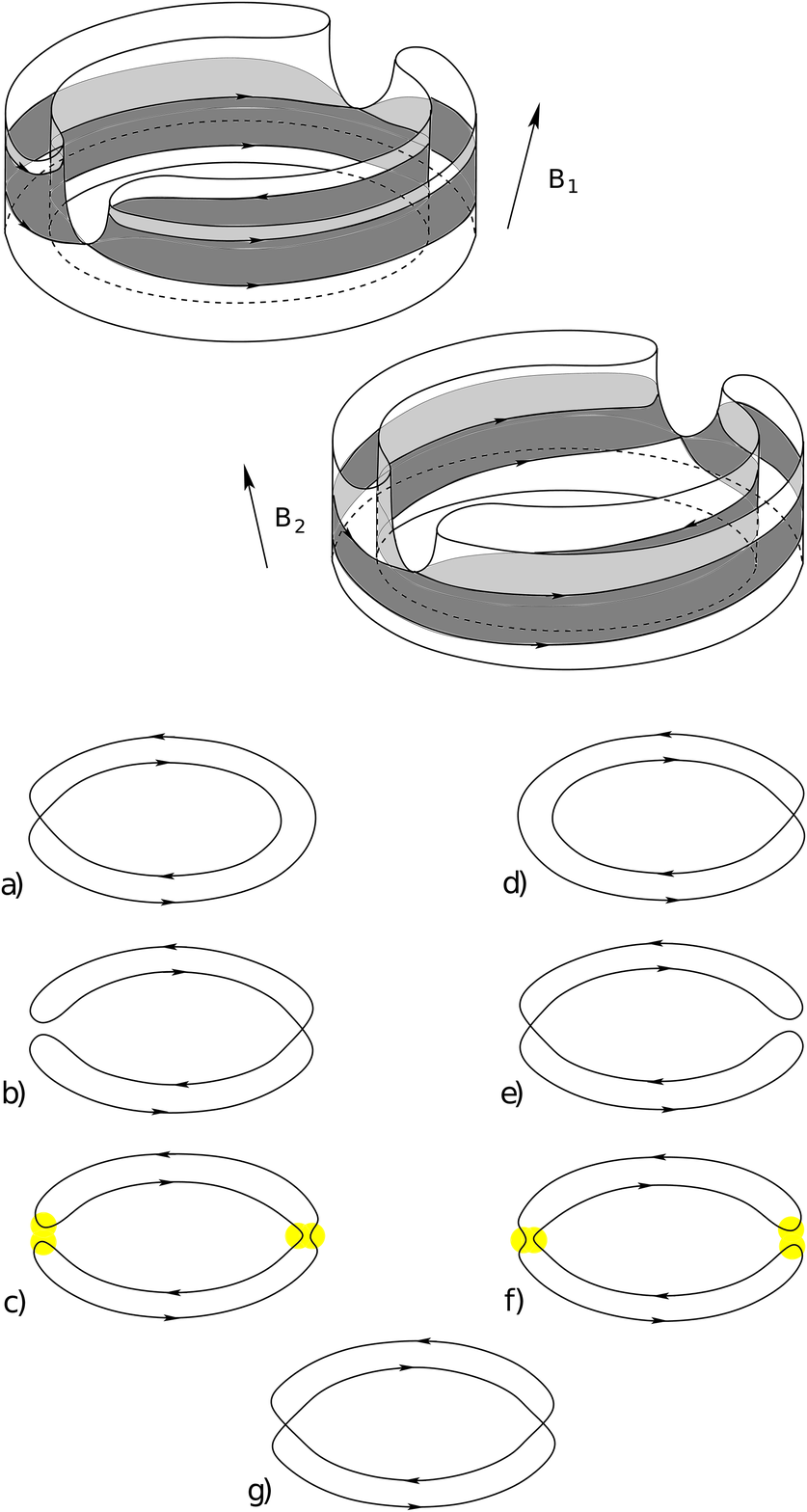}
\end{center}
\caption{Reconstruction of the structure of (\ref{MFSyst}) 
without central symmetry. The trajectories of the extreme 
area on cylinders of low height are absent both before and 
after the reconstruction. The trajectories of the minimum 
orbital period on opposite sides of the reconstruction are of 
the same type (electron or hole).}
\label{Fig16}
\end{figure}

 The reconstruction shown in Fig. \ref{Fig16}, differs from 
all other reconstructions discussed above. Namely, here each 
of the extreme trajectories has four ``deceleration sections'' 
(on each side of the reconstruction boundary). When crossing 
the corresponding boundary of the reconstruction, the full 
addition to the orbital period due to the ``deceleration segments'' 
does not change and remains equal to 
$\, 2 \left( T_{1} (\alpha) + T_{2} (\alpha) \right) \, $.
It is easy to see that this property makes it possible to 
unambiguously identify this reconstruction among all the 
reconstructions of the first group.

 The remaining types of reconstructions of the system 
(\ref{MFSyst}), shown in Fig. \ref{Fig17} - \ref{Fig21},
on the contrary, have the property that they contain 
trajectories of extreme area on the cylinders of small height, 
both before and after the reconstruction. Together with the 
reconstruction shown in Fig. \ref{Fig10}, they form the second 
class of reconstructions, complementing the class of 
reconstructions shown in Fig. \ref{Fig11} - \ref{Fig16}. 
All these reconstructions are experimentally easily 
distinguishable from those of the first class, since, 
along with a jump in the picture of classical oscillations, 
they also have a jump in the picture of quantum oscillations 
(an abrupt replacement of some oscillation terms by others). 
For the experimental distinction between the reconstructions 
of this class, one can also use the above (and also other) 
features of the oscillatory picture in observing classical 
oscillations. But, of course, these reconstructions also differ 
from each other by the features of changes in the picture of 
quantum oscillations, which we would like to dwell on below.

\begin{figure}[t]
\begin{center}
\includegraphics[width=0.9\linewidth]{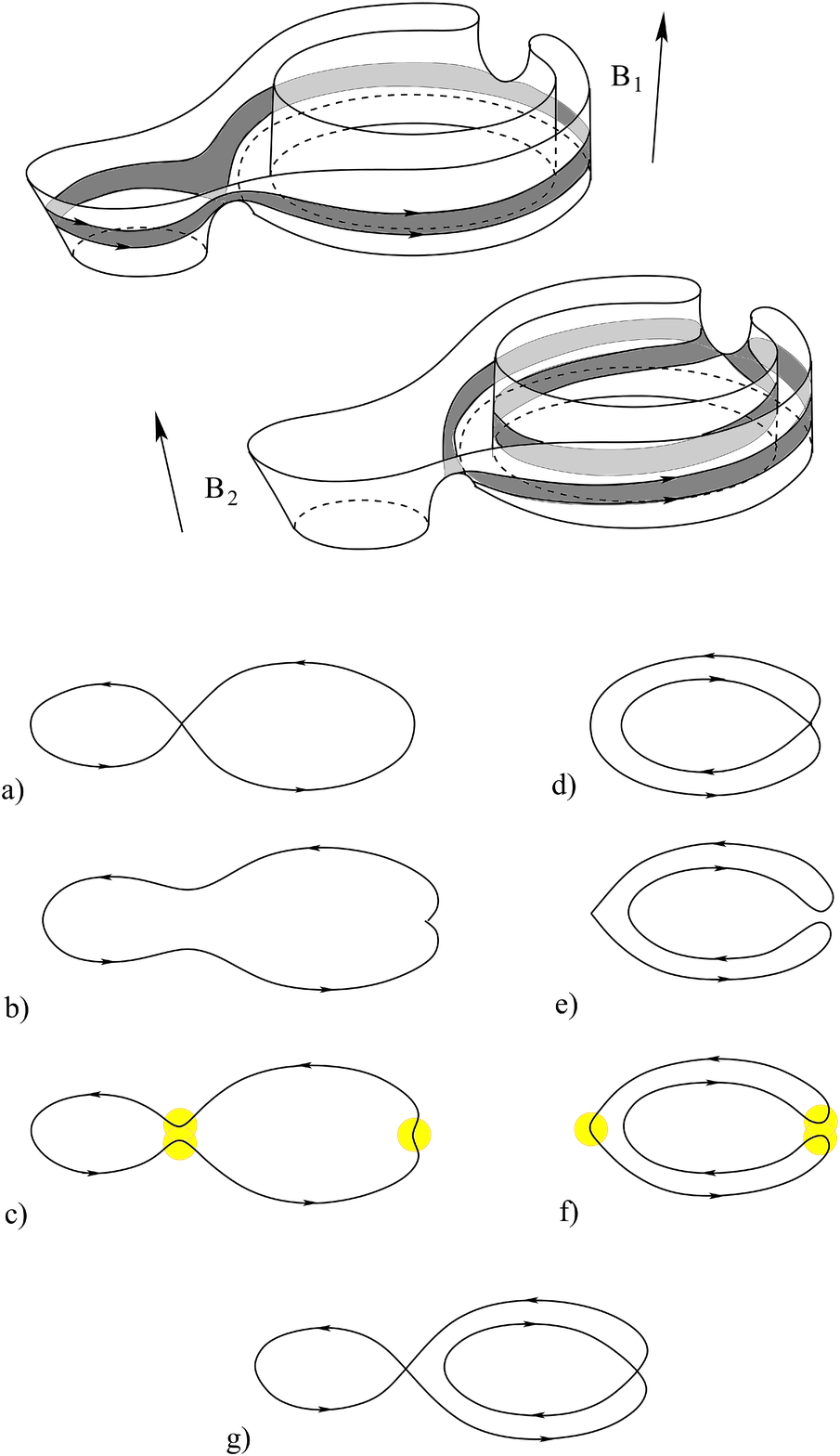}
\end{center}
\caption{Reconstruction of the structure of (\ref{MFSyst}) 
without central symmetry. Extreme area trajectories on low-height 
cylinders are present both before the reconstruction 
(the maximum area trajectory) and after the reconstruction 
(the minimum area trajectory). The area of the extreme trajectory 
before the reconstruction is always larger than the area of the 
extreme trajectory after the reconstruction. The extreme 
trajectories on both sides of the reconstruction are of the 
same (electron or hole) type.}
\label{Fig17}
\end{figure}

\begin{figure}[t]
\begin{center}
\includegraphics[width=0.9\linewidth]{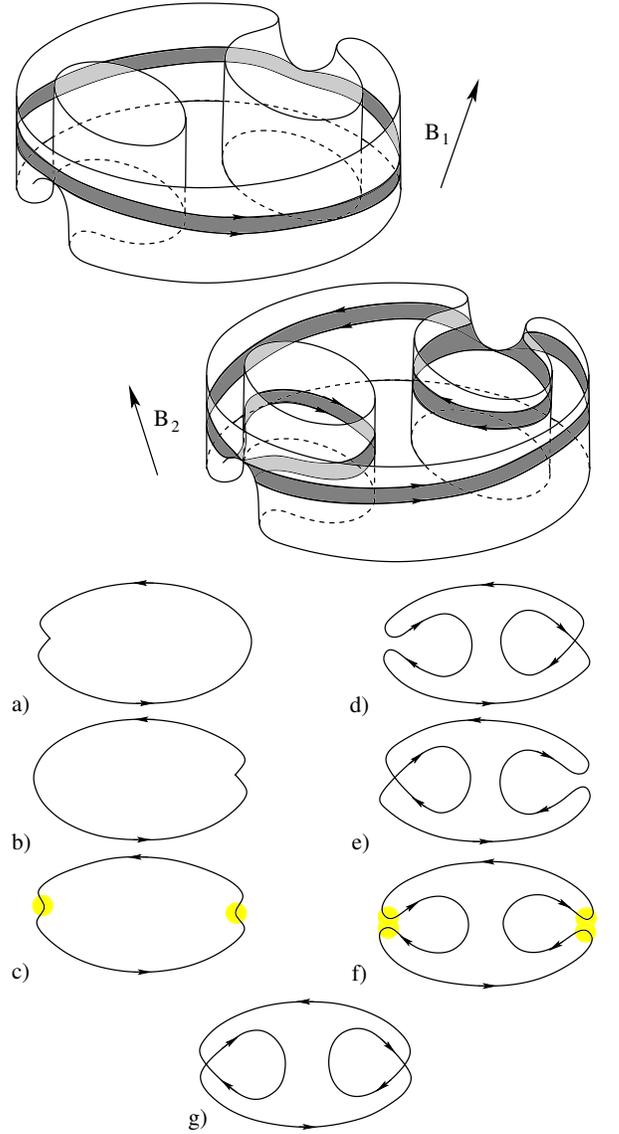}
\end{center}
\caption{Reconstruction of the structure of (\ref{MFSyst}), 
which can have central symmetry. Extreme area trajectories 
on low-height cylinders are present both before the 
reconstruction (the maximum area trajectory) and after the 
reconstruction (the minimum area trajectory). The area of the 
extreme trajectory before the reconstruction is always larger 
than the area of the extreme trajectory after the reconstruction. 
The extreme trajectories on both sides of the reconstruction 
are of the same (electron or hole) type.}
\label{Fig18}
\end{figure}

 Note right away that the reconstructions, shown in 
Fig. \ref{Fig10}, \ref{Fig17} - \ref{Fig18}, differ from the 
reconstructions shown in Fig. \ref{Fig19} - \ref{Fig21}.
Namely, for all reconstructions shown in Fig. \ref{Fig10}, 
\ref{Fig17} - \ref{Fig18}, the extreme area trajectories 
have a minimum area on one side of a reconstruction and a 
maximum area on the other side. In our case, the minimality 
(maximality) of the trajectory area means that the trajectory 
area increase (decrease) when approaching the bases of the 
corresponding cylinder of closed trajectories for a fixed 
direction of $\, {\bf B} \, $. This, as we have already said, 
is due to the presence of singular points at the bases of 
such cylinders. In fact, for the same reason, the same 
increase (decrease) in the minimum (maximum) trajectory 
area (with an infinitely increasing derivative) occurs 
when the direction of the magnetic field approaches the 
boundary of a reconstruction of the structure of (\ref{MFSyst}) 
(and a decrease in the height of the corresponding cylinder 
of closed trajectories to zero). This circumstance makes it 
possible to easily identify the trajectories of the minimum 
and maximum area described by us here, and, in particular, 
to distinguish experimentally the reconstructions shown in
Fig. \ref{Fig10}, \ref{Fig17} - \ref{Fig18}, from the 
reconstructions shown in Fig. \ref{Fig19} - \ref{Fig21}.
As also seen from Fig. \ref{Fig10}, \ref{Fig17} - \ref{Fig18}, 
in all these cases, the maximum area trajectories have 
a larger area than the minimum area trajectories.

 In addition to the above circumstance, it can also be noted 
that in all the reconstructions shown in Fig. \ref{Fig10}, 
\ref{Fig17} - \ref{Fig18}, the trajectories of the extreme 
area are of the same type (electron or hole) before and after 
the reconstruction. This circumstance can also be easily 
established experimentally from the behavior of quantum 
oscillations of transverse (Hall conductivity), which also 
makes it possible to distinguish these reconstructions from 
those shown in Fig. \ref{Fig19} - \ref{Fig21}. 

 As for the distinguishing the reconstructions shown in
Fig. \ref{Fig10}, \ref{Fig17} - \ref{Fig18}, from each other, 
then here, as above, we can immediately note a special feature 
of the reconstruction in Fig. \ref{Fig17}, consisting in the 
presence of three sections of deceleration on the trajectories 
of cylinders of low height, both on one side and on the other 
side of the reconstruction. This circumstance makes it possible 
to immediately distinguish the reconstruction in Fig. \ref{Fig17} 
from the other two, for example, by measuring the orbital period 
along the trajectory when observing classical oscillations or the 
temperature dependence of quantum oscillations (replacing of
$\, \Delta T (\alpha) = 2 T_{1} (\alpha) + T_{2} (\alpha) \, $
by $\, T_{1} (\alpha) + 2 T_{2} (\alpha) $). But, actually, 
this fact can be easily established also by simple observation 
of quantum oscillations from the behavior of the area 
$\, S (\alpha) \, $ of the extreme trajectory near the 
transition boundary, where its main dependence on 
$\, \alpha \, $ is determined precisely by the approache 
to  singular points of the system (\ref{MFSyst}). 
To distinguish the reconstructions shown in Fig. \ref{Fig10} 
and Fig. \ref{Fig18}, it is possible, for example, to investigate 
the possibility of falling of a deceleration section into the skin 
layer at the sample boundary (\cite{OsobCycle}) in observing
the classical cyclotron resonance (it is possible on one side 
of a reconstruction for the reconstruction in Fig. \ref{Fig10} 
and it is impossible for the reconstruction in Fig. \Ref{Fig18} 
due to the peculiarities of the geometry of the trajectories).

\begin{figure}[t]
\begin{center}
\includegraphics[width=0.9\linewidth]{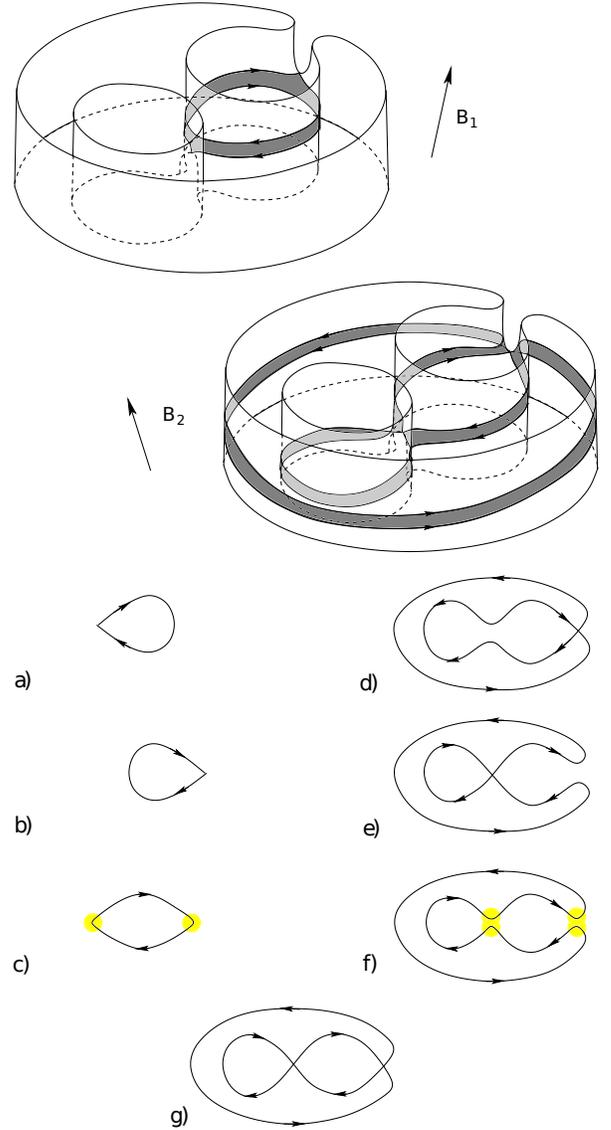}
\end{center}
\caption{Reconstruction of the structure of (\ref{MFSyst}) 
without central symmetry. Extreme-area trajectories on low-height 
cylinders are present both before the reconstruction 
(the trajectory of the minimum area) and after the reconstruction 
(the trajectory of the minimum area). The extreme trajectories on 
opposite sides of the reconstruction are of different types 
(electronic, on the one side of the reconstruction, and hole-type, 
on the other).}
\label{Fig19}
\end{figure}

\begin{figure}[t]
\begin{center}
\includegraphics[width=0.9\linewidth]{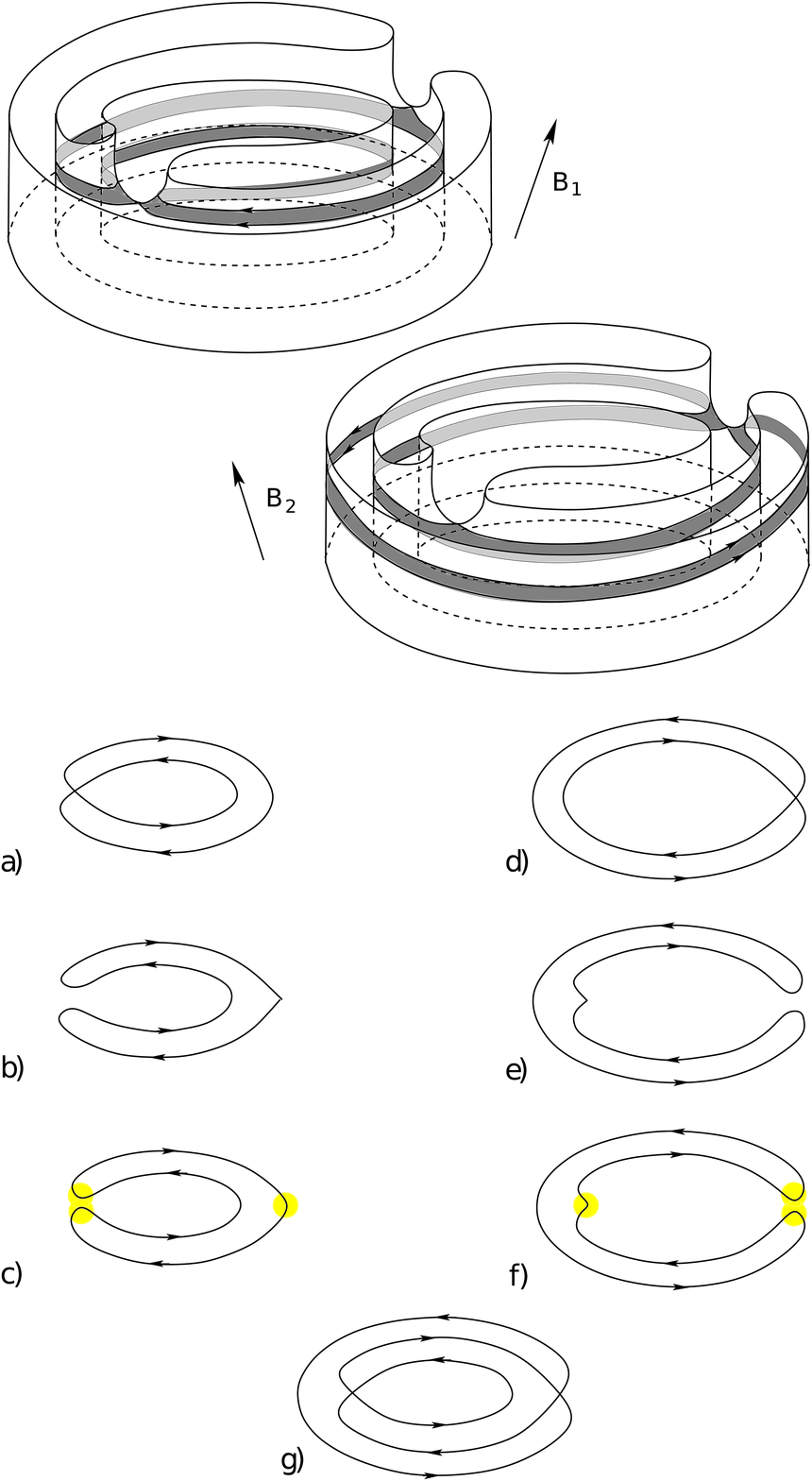}
\end{center}
\caption{Reconstruction of the structure of (\ref{MFSyst}) 
without central symmetry. Extreme-area trajectories on low-height 
cylinders are present both before the reconstruction 
(the trajectory of the minimum area) and after the reconstruction 
(the trajectory of the minimum area). The extreme trajectories on 
opposite sides of the reconstruction are of different types 
(electronic, on the one side of the reconstruction, and hole-type, 
on the other).}
\label{Fig20}
\end{figure}

 For the reconstrictions shown in Fig. \ref{Fig19} - \ref{Fig20}, 
extreme trajectories have a minimum area on both the disappearing 
and arising cylinders of closed trajectories. In both of these 
reconstructions, the extreme trajectories have different types 
(electronic and hole) on opposite sides of a reconstruction. 
Both these circumstances can be easily established by observing 
quantum oscillations of various types (the De Haas - Van Alphen effect, 
the Shubnikov - De Haas effect) and distinguish the reconstructions 
in Fig. \ref{Fig19} - \ref{Fig20} from all other reconstructions. 
The difference between the reconstructions shown in Fig. \ref{Fig19} 
and Fig. \ref{Fig20}, consists, for example, in the number of 
deceleration sections on the corresponding trajectories before 
and after the reconstruction. As we have already seen above, 
this difference can also be easily established by observing both 
classical and quantum oscillations.

\begin{figure}[t]
\begin{center}
\includegraphics[width=0.9\linewidth]{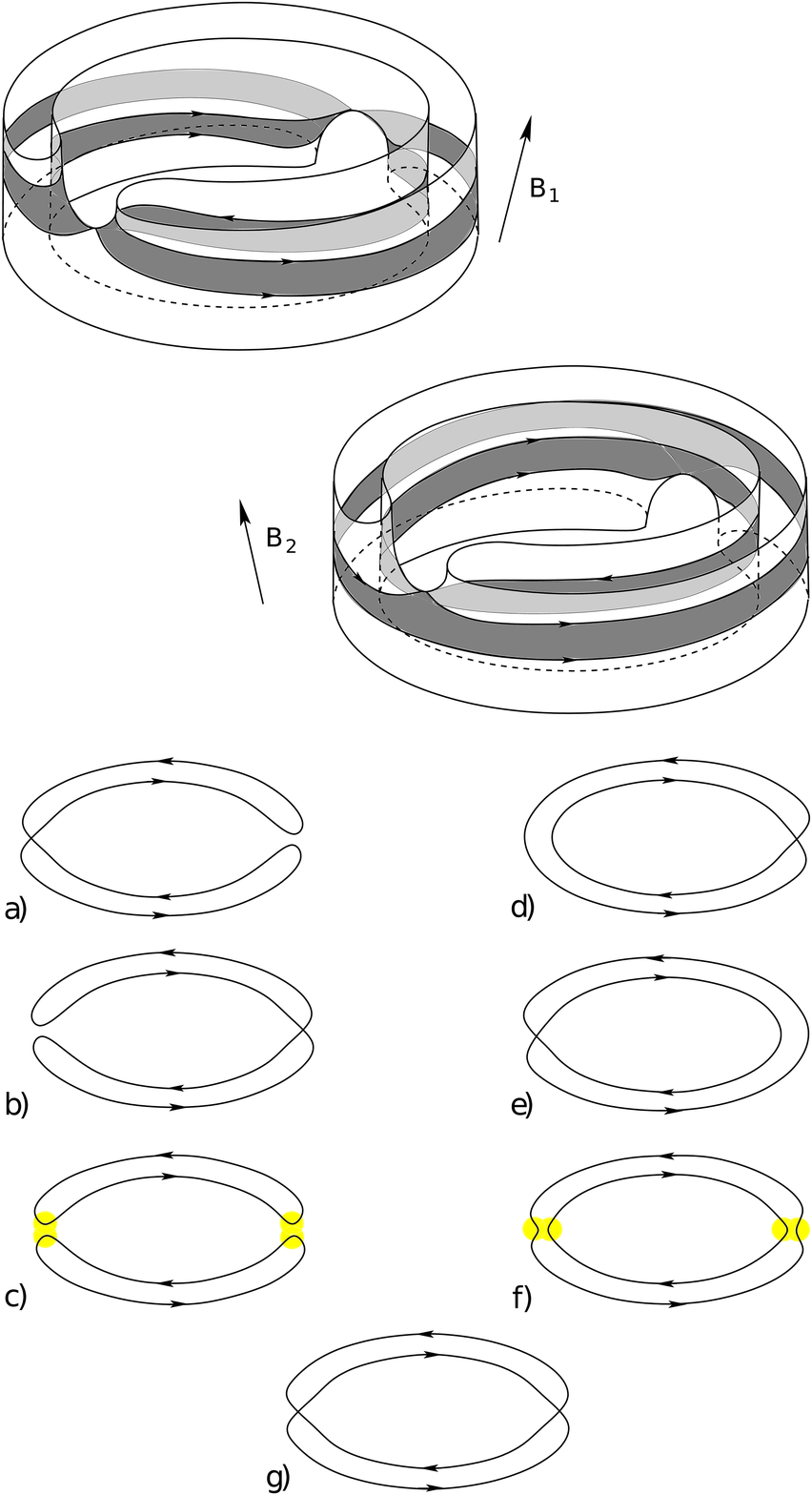}
\end{center}
\caption{Reconstruction of the structure of (\ref{MFSyst}), 
which can have central symmetry. Extreme area trajectories 
on cylinders of low height are present both before the 
reconstruction (a pair of trajectories of the minimum area) 
and after the reconstruction (one trajectory of the minimum 
area and one trajectory of the maximum area). A pair of extreme 
trajectories on one side of the reconstruction are of the same 
type. After the reconstruction, a pair of extreme trajectories 
of opposite types appear, and the type of the trajectory of 
a larger area coincides with the type of extreme trajectories 
before the reconstruction.}
\label{Fig21}
\end{figure}

 The reconstruction shown in Fig. \ref{Fig21}, as it is easy 
to see, differs in many aspects from all the reconstructions
we have considered earlier. In this reconstruction, a pair of 
extreme trajectories disappear and appear at once. On the one 
side of the reconstruction, both trajectories are of the same 
type (electronic or hole), while on the other side, a pair of 
trajectories of different types appears. This circumstance 
can be determined, in particular, by observing quantum 
oscillations of conductivity, which allows one to immediately 
identify this reconstruction experimentally. We also note here 
that the reconstruction shown in Fig. \ref{Fig21} may have 
central symmetry, which, in fact, almost always implies such 
symmetry in real situations.

\vspace{1mm}

 All our considerations above was carried out without taking 
into account the electron spin. It is easy to see that taking 
into account the spin states leads to the splitting of each of 
the oscillatory terms into two, in accordance with the direction 
of the spin along or against the direction of $\, {\bf B} $. 
In addition, in our consideration, we also did not take into 
account the effect of the Berry phase, which can manifest 
itself in materials with no inversion center or symmetry 
with respect to time reversal. The appearance of nonzero Berry 
curvature in such materials can also lead to a number of 
interesting effects in the situation under consideration.

\vspace{1mm}

 In conclusion, we would like to mention here also the 
phenomenon of magnetic breakdown, which can be observed 
in our situation. As is well known, the phenomenon 
of (intraband) magnetic breakdown is observed 
in many substances in sufficiently strong magnetic fields and, 
in particular, can lead to many interesting effects, arising 
on trajectories of system (\ref{MFSyst}) of different geometry 
(see, for example
\cite{etm,Zilberman1,Zilberman2,Zilberman3,Azbel,Slutskin1967,
AlexsandradinGlazman}).
Thus, it can be expected that the occurrence of a magnetic 
breakdown on the special extreme trajectories described above 
should also lead to interesting phenomena, in particular, to 
significantly affect the quantization of electronic levels 
for trajectories of the extreme area. It must be said, however, 
that the magnetic breakdown on the described trajectories 
can occur only at rather large values of $\, B \, $ and only 
when the direction of the magnetic field approaches the boundary 
of a reconstruction of the topological structure of system 
(\ref{MFSyst}) very closely (see \cite{OsobCycle}). As a consequence 
of this, a clear picture of the corresponding effects should be 
observed only in very precise experiments, which make it possible 
to specify the directions of the magnetic field with a very high 
accuracy (and can be noted only as a certain blurring of the 
sharp changes in the oscillation picture described above 
in a very narrow region near the reconstruction boundary 
at a lower resolution). In this situation, certainly, 
the dependence of the picture (especially of quantum) 
of oscillations on the topological type of a reconstruction 
under conditions of developed magnetic breakdown is also of 
great interest.

\section{Conclusion}

 The paper considers the features of oscillatory phenomena 
in metals near the boundaries of reconstruction of the 
topological structure of the system describing the adiabatic 
dynamics of quasiparticles on complex Fermi surfaces. 
Each elementary reconstruction of such a structure is 
associated with a change in the picture of closed trajectories 
on the Fermi surface, which consists in the disappearance of 
some of the cylinders of closed trajectories and the 
appearance of new ones. Each of these reconstructions has its 
own topological structure, and there are a finite number of 
topological types of such reconstructions. The most important 
circumstance in each of the reconstructions is the disappearance 
of a part of closed trajectories with extreme values of the area 
or the orbital period, which leads to sharp observable changes 
in the picture of oscillatory phenomena during the reconstruction.
The features of such changes are directly related to the geometry 
of disappearing and arising extreme trajectories, which is 
determined by the topological type of a reconstruction. 
The paper presents a detailed comparison of the features of the 
change in the picture of classical and quantum oscillations 
at the moment of a reconstruction with its topological type 
and proposes methods for identifying topological types based 
on these features. The proposed methods, in our opinion, 
can be rather useful in studying the geometry of complex 
Fermi surfaces using classical or quantum oscillations 
in strong magnetic fields.

\vspace{5mm}

This work is supported by the Russian Science Foundation 
under grant 21-11-00331.

\end{document}